\documentclass[11pt,twoside]{article}
\usepackage{asp2014}

\aspSuppressVolSlug
\resetcounters

\begin{document}

\title{Antennas and Receivers in Radio Interferometry}
\author{Todd R. Hunter$^1$ and Peter J. Napier$^2$}
\affil{$^1$NRAO, Charlottesville, VA, USA; \email{thunter@nrao.edu}}
\affil{$^2$NRAO, Socorro, VA, USA; \email{pnapier@nrao.edu}}

\paperauthor{Todd~R.~Hunter}{thunter@nrao.edu}{0000-0001-6492-0090}{NRAO}{}{Charlottesville}{VA}{22903}{USA}
\paperauthor{Peter~J.~Napier}{pnapier@nrao.edu}{}{NRAO}{}{Socorro}{VA}{87801}{USA}

\begin{abstract}

  The primary antenna elements and receivers are two of the most
  important components in a synthesis telescope. Together they are
  responsible for locking onto an astronomical source in both
  direction and frequency, capturing its radiation, and converting it
  into signals suitable for digitization and correlation.  The
  properties and performance of antennas and receivers can affect the
  quality of the synthesized images in a number of fundamental ways.
  In this lecture, their most relevant design and performance
  parameters are reviewed, with emphasis on the current ALMA and VLA
  systems.  We discuss in detail the shape of the primary beam and the
  components of aperture efficiency, and we present the basics of
  holography, pointing, and servo control.  On receivers, we outline
  the use of amplifiers and mixers both in the cryogenic front-end and
  in the room temperature back-end signal path.  The essential
  properties of precision local oscillators (LOs), phase lock loops
  (PLLs), and LO modulation techniques are also described. We provide
  a demonstration of the method used during ALMA observations to
  measure the receiver and system sensitivity as a function of
  frequency.  Finally, we offer a brief derivation and numerical
  simulation of the radiometer equation.

\end{abstract}

\section{Introduction}

In this lecture, we present the most important aspects of the antenna
and receiver components of synthesis telescopes.  Due to the increased
breadth of material, we cannot cover all of the topics contained in
the previous version of the summer school chapter on antennas
\citep{Napier99}, in particular the section on antenna polarization
properties.  Instead, we review the basics of antennas while adding
new details of interest to astronomers on the Atacama Large
Millimeter/submillimeter Array (ALMA) and the Karl G. Jansky Very
Large Array (VLA) dish reflectors.  We follow with an overview of the
heterodyne receiver systems and the receiver calibration techniques in
use at these telescopes.

Figure~\ref{fig1} shows a simple block diagram of the major components
required in an synthesis telescope.  The role of the primary antenna
elements of an interferometer is much the same as in any single
element telescope: to track and capture radiation from a celestial
object over a broad collecting area and focus and couple this signal
into a receiver so that it can be detected, digitized and analyzed.
At the output of the receiver feed, the signal is at the radio, or
sky, frequency $\nu_{\rm RF}$, typically with a significant bandwidth
$\Delta\nu$.  The signal undergoes frequency translations and
filtering as it propagates through the electronics system.  In
synthesis telescopes of recent design, the analog signal processing
and digitization systems are located in the antennas, with the
resulting digital data transmitted over fiber optic cables to the
correlator building.  In general, the receiver, intermediate
frequency, transmission cables, LO, and baseband portions of the
electronics system all have the requirement of good amplitude and
phase stability. These requirements and others such as stable bandpass
shape, low spurious signal generation, and good signal isolation are
discussed in \citet{Thompson07} and in the ALMA and EVLA Memo series.

\articlefigure{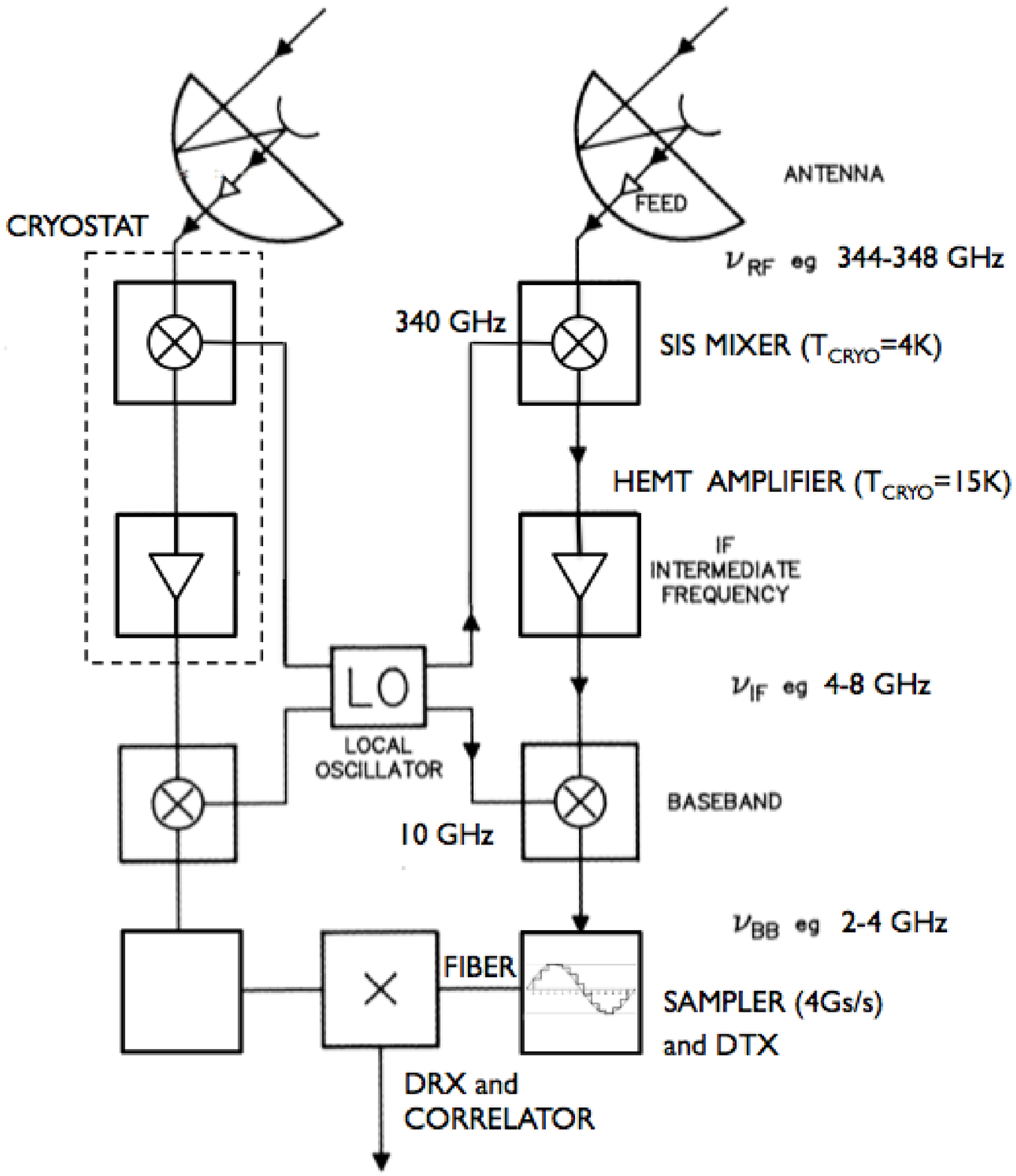}{fig1}{A simplified block diagram of the
  electronic equipment used to produce the correlation from one
  antenna pair in a synthesis telescope. The signal frequencies given
  as examples at various points through the electronics chain are
  typical of ALMA observing in Band 7 (LO1=340~GHz), but only one of
  the four dual-polarization basebands and samplers is shown
  (LO2=10~GHz).  The digital transmitter (DTX) and digital receiver
  (DRX) relay the sampled data from the antenna to the central
  building via optical fiber.  For lower frequency interferometers
  (like the VLA), the order of the first mixer and amplifier is
  swapped and the mixer is a room temperature device.}

\section{Antennas}

Historically, a great variety of antenna types have been employed in
synthesis telescopes \citep[see the list in][Table 1]{Napier83}.  In
all cases, the diffraction beam of the primary antenna defines the
solid angle over which an interferometer is sensitive, and is called
the {\it primary beam}.  The details of this angular response pattern,
including beam shape, sidelobe level, and polarization purity, as well
as how accurately it can track the target are important, and will
directly affect the observed data.


\subsection{Antenna types}

The three major categories of antennas used in radio astronomy
include: simple dipole antennas, horn antennas, and parabolic
reflecting antennas.  Dipole antennas provide the widest field
response but at low gain, meaning that large arrays of hundreds or
thousands of them are necessary to form beams with any reasonable
level of resolution and sensitivity.  They are typically used at
wavelengths longer than 1~m, such as in the Long Wavelength Array
\citep[LWA,][]{Ellingson09}.  Horn antennas provide the most
well-controlled beam shape and uniformity of response vs. frequency.
Indeed, a hybrid of the horn and parabolic reflecting antenna types,
the Crawford Hill horn-reflector \citep{Crawford61}, yielded the first
detection of the cosmic microwave background \citep{Penzias65}.  Horn
antennas have been combined into small interferometric arrays built on
tracking platforms such as the Degree Angular Scale Interferometer
\citep{Halverson02}.  However, horn antennas are not practical when a
large collecting area is required because large single elements would
be long, heavy, and difficult to arrange into a compact configuration.
The reflecting dish antenna provides both good sensitivity and beam
performance in its single element form, while being amenable to
arrangement into reconfigurable interferometric arrays.
In order to access a wide frequency range, many different receiver
bands must be arranged with some mechanism to share the focal
plane. However, this issue has been solved by a variety of
approaches. For example, at the Green Bank Telescope
  \citep[GBT,][]{Prestage09}, up to eight receivers are mounted on a
  circular carriage which can rotate the selected receiver onto the
  focal axis.  Because ALMA, VLA and many other major
interferometers employ symmetric dish antennas with circular
apertures, we will concentrate the rest of this section on this style
of antenna.

\subsection{Design of reflector antennas}

Since the mid-1960's, reflector antennas have been designed using the
principle of homology \citep{vonHoerner67}.  Rather than trying to
build a structure to resist the deformation associated with changes in
orientation, a homologous design responds to the changes by allowing
the surface to perturb from one parabola to another.  This change can
then be compensated simply by applying a calibrated, concomitant
motion (i.e. focus) of the subreflector.  Further discussion on
homology can be found in \citet{Baars07}, which is an excellent
reference on performance measurements techniques for parabolic
reflector antennas.  Structural engineering of antennas is discussed
in \citet{Levy96}.  To summarize in a single sentence, the typical
modern antenna presents a thin aluminum reflecting surface composed of
dozens to hundreds of molded or machined segments supported by a space
frame backup structure (BUS) composed of carbon fiber reinforced (CFR)
tubes and/or steel members and fasteners.  These components promote
high surface efficiency while offering some immunity to thermal
deformation.  The two major choices when designing a reflector antenna
for use in a synthesis array are the choice of mount and the choice of
optics.
 
For radio astronomy dishes, the alt-azimuth mount is the most
prevalent in use today.  Its advantages are its simplicity and the
fact that gravity always acts on the reflector in the same plane,
easing the challenge of a homologous design. The major disadvantage of
this mount is that, as the antenna tracks, the aperture (and hence the
primary beam) rotates with respect to the source, around the primary
optical axis \citep{Thompson07}.  If the source size is significant
compared to the beam size, and if the beam is not circularly
symmetric, this rotation will cause the apparent brightness
distribution to vary.  Since aperture blockage usually makes the beam
sidelobe pattern non-circularly symmetric, and the antenna
instrumental polarization is not circularly symmetric, the dynamic
range of total intensity images of very large sources and polarization
images of extended sources will be limited by this effect. Observers
of extended sources need to consider this effect when judging the
fidelity of subtle features in the images of these sources.  A minor
disadvantage of the alt-az mount is that sources passing close to the
zenith cannot be tracked well due to the high rates of azimuth
rotation needed.  Technically, the sidereal azimuth rate exceeds the
(relatively slow) slew rate of the VLA and GBT ($\sim40\deg$/minute)
only for elevations $>89.67\deg$.  However, many antenna servos are
not necessarily designed for smooth tracking at high rates, so errors
may be larger (\S~\ref{tracking}).  Often of greater concern is the
typically reduced accuracy of the pointing model at elevations
$>80\deg$ (\S~\ref{pointing}).

There are a variety of optical systems that can be used to feed a
large radio reflector \citep[e.g.][]{Rudge82}.  Figure~\ref{fig2}
shows the major types of feed systems that are in use on current radio
telescope reflectors.  The {\it prime focus system}, as in the
Westerbork Synthesis Radio Telescope \citep[WSRT,][]{Baars74} and the
Giant Meterwave Radio Telescope \citep[GMRT,][]{Ananthakrishnan95},
has the advantage that it can be used over the full frequency range of
the reflector, including the lowest frequencies where secondary focus
feeds become impractically large. The disadvantages of the prime focus
are that space for, and access to, the feed and receiver is restricted
and spillover noise from the ground decreases sensitivity.  All of the
{\it multiple reflector systems} (Figure~\ref{fig2}(b)--(f)) have the
advantage of more space, easier access to the feed and receiver, and
reduced noise pickup from the ground.  In addition, the primary and
secondary reflectors can be shaped to provide more uniform
illumination in the main reflector aperture, as described in
\S~\ref{shaped}.  The {\it off-axis Cassegrain} (e.g., VLA, VLBA,
ALMA) is particularly suitable for synthesis telescopes needing
frequency flexibility, because many feeds can be located in a circle
around the main reflector axis.  Changing frequency simply requires
either a rotation of the asymmetric subreflector around this axis, as
in the VLA \citep{Napier83} and VLBA \citep{Napier94}, or by adjusting
the pointing of the primary mirror as in ALMA \citep{Hills10}.  The
disadvantage of this geometry is that the asymmetry degrades
polarization performance.  The {\it Nasmyth geometry} \citep[e.g., the
10.4~m Leighton dishes of the Combined Array for Millimeter Astronomy
(CARMA),][]{Woody04} provides a receiver cabin external to the antenna
structure, whilst the {\it bent Nasmyth geometry}\/ (e.g.,
Submillimeter Array (SMA)) minimizes disturbances to the receivers
because they (along with the final three mirrors) do not tilt in
elevation \citep{Paine94}.  The bent Nasmyth geometry provides maximum
convenience for service access, even during observations.  Finally,
the {\it dual offset Gregorian} (e.g., GBT) has no blockage and thus
delivers a circularly symmetric beam with low sidelobes which is
particularly important for Galactic H~I observations
\citep{Boothroyd11}.  This characteristic makes it an attractive
choice for wide field-of-view synthesis telescopes, but the increased
complexity of reflector panel tooling and subreflector support
structure leads to increased cost.

\articlefigure{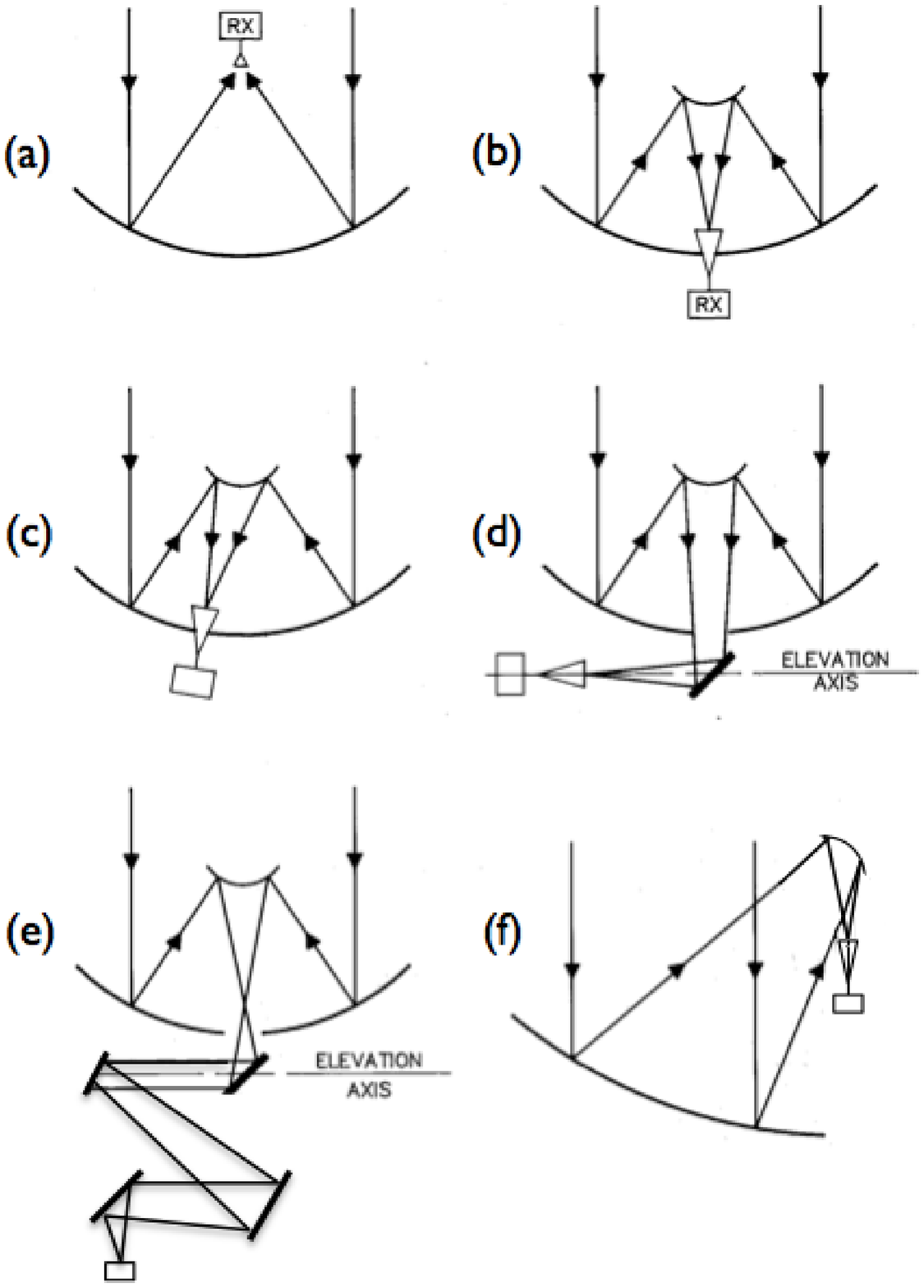}{fig2}{Optical systems for radio telescope reflectors.
{\bf (a)}\ Prime focus,
{\bf (b)}\ Cassegrain,
{\bf (c)}\ Off-axis Cassegrain,
{\bf (d)}\ Nasmyth,
{\bf (e)}\ Bent Nasmyth,
{\bf (f)}\ Dual offset Gregorian.}

\subsection{Antenna primary beam shape}  

\label{shaped}

As described in \citet{Napier99}, there is a Fourier transform
relationship between the complex voltage distribution of the electric
field, $f(u,v)$, in the aperture of an antenna and the corresponding
complex far-field voltage radiation pattern, $F(l,m)$ of the antenna
\citep[see also][\S 6--8]{Kraus86}.  In both domains, the power
pattern is the square of the absolute magnitude of the voltage
pattern.  The form of $f(u,v)$ for an antenna is determined by the way
in which the antenna feed illuminates the aperture. In general, the
more that $f(u,v)$ is tapered at the edge of the aperture, the lower
will be the aperture efficiency and the sidelobe response, and the
broader the main beam. Calculations for a variety of $f(u,v)$, and
their corresponding $F(l,m)$, can be found in antenna textbooks
\citep[e.g.][Chapter 4]{Baars07}.  Figure~\ref{fourierpair} shows
one-dimensional cuts through $f(u,v)$ and $|F(l,m)|^2$ for uniform and
tapered illumination patterns.

\articlefigure{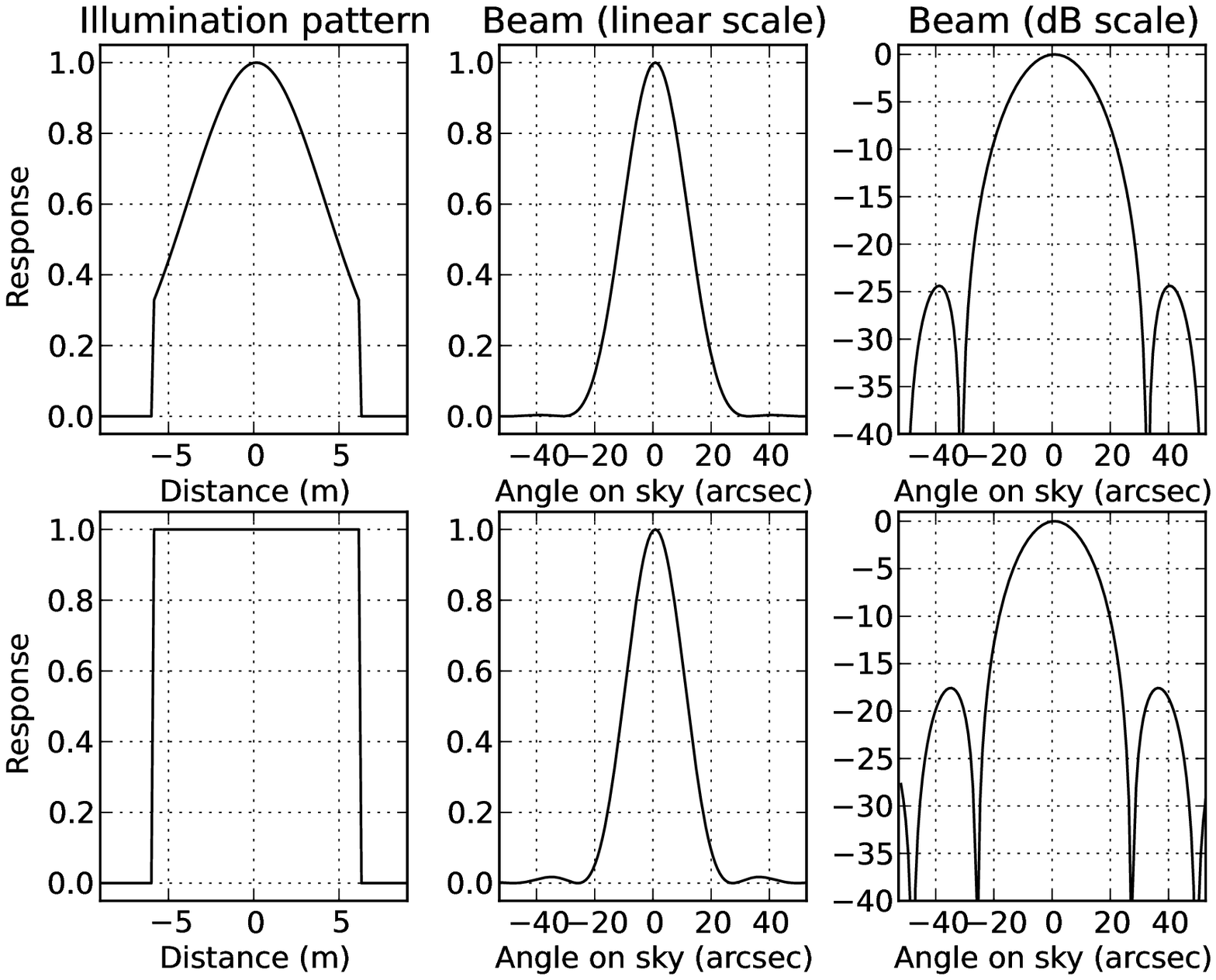}{fourierpair}{These plots are
  one-dimensional cuts through two-dimensional images simulating an
  (unblocked) 12~m diameter antenna observing at 230~GHz, showing the
  relationship between an antenna's aperture voltage pattern $f(u)$
  (left column) and its far-field radiation power pattern $|F(l)|^2$
  (right columns).  The top row shows the case of a Gaussian edge
  taper (-10~dB in power) while the bottom row shows what would happen
  with uniform illumination.  In general, both quantities are complex
  but only the amplitudes are shown here. Note the difference in beam
  width and sidelobe level.}
 
In order to maximize sensitivity (at the expense of higher sidelobes),
the VLA antennas were designed to have a nearly uniform illumination
($f(u,v)={\rm constant}$) over the whole aperture, except where the
aperture is blocked by the subreflector and its support
struts. Because efficient receiver feedhorns have a tapered response,
achieving this uniform illumination required mathematically perturbing
the primary and secondary mirror surfaces into complementary
``shaped'' surfaces \citep{Williams65}.  In other words, the VLA antennas are {\bf not}
the classical Cassegrain combination of paraboloidal primary with
hyperboloidal secondary.  While the primary differs from a paraboloid
by only 1~cm rms \citep{Napier83}, recent optical modeling employed a
polynomial of order 13 to accurately represent the surface
\citep{Srikanth05}.  The VLBA antennas are similarly shaped to provide
uniform illumination out to 95\% of the dish radius, then tapered to
-15~dB at the edge \citep{Napier94}.

With uniform illumination, for a circularly symmetric aperture of
diameter $D$, the beam pattern takes the form $F(u)=J_1(\pi Du)/u$,
which has the following properties: first sidelobe level $=-17.57$~dB
(i.e. $10^{-1.757}$ compared to the peak), half power beam width
HPBW~$=1.028\lambda/D$, and the radius of the first null $=1.22\lambda
/D$.  These values are in good agreement with measured beam parameters
for the VLA 25~m diameter reflector, except for the first sidelobe
level (about -16~dB), which is increased from theory by the aperture
blockage. The VLA beam patterns in the various bands are characterized
by sixth-order polynomial functions in the AIPS software package, and
analytically by an Airy pattern (truncated at the 10\% level) in the
CASA software package.  More accurate patterns are being added to
CASA. Some disadvantages of shaped Cassegrain geometries, which do not
usually preclude their use for synthesis telescopes, include
compromised prime focus operation above a frequency of about 1~GHz
because of the shaped main reflector (the VLBA 600~MHz system suffers
5\% loss due to this effect), and very bad beam degradation if the
feed is moved away from the secondary focal point. This latter problem
can limit their use in synthesis arrays designed to obtain very wide
fields of view by using focal plane arrays (FPAs).  Note that the
Apertif FPA system \citep{Oosterloo09}, which has a 8\deg$^2$ field of
view at 1.4~GHz, operates at prime focus on the WSRT whose 25~m
antennas are paraboloidal in shape \citep{Baars74} and thus do not
suffer from this complication.

In contrast, ALMA antennas were designed to have a tapered
illumination pattern because it provides reduced sidelobes which
promotes better single dish imaging performance, a required capability
of ALMA.  In addition, the classical Cassegrain geometry provides good
performance over a much larger area of the focal plane, which the
stationary ALMA receivers must share.  By specification, the taper of
the receiver feeds at the edge of the subreflector is $-12$~dB in the
Gaussian beam approximation, which equates to $-10$~dB in the physical
optics analysis \citep{Rudolf07}.  A quadratic taper of $-10$~dB
(i.e. the power at the edge of the dish is 10\%\/ of the peak)
corresponds to a HPBW of $1.137\lambda/D$ \citep[][Eq 4.13]{Baars07}.
The central obstruction of 0.75~m on the 12~m antennas produces a
further $\sim0.5$\%\/ reduction in the beam pattern
\citep{Schroeder87} to a final HPBW of $1.13\lambda$/D.  The
theoretical peak of the first sidelobe is $-24$~dB for an unblocked
aperture. The effect of a central blockage is to increase the
odd-numbered sidelobes by a few dB while similarly decreasing the
even-numbered sidelobes.  Currently in CASA, the ALMA beam pattern is
an Airy pattern scaled to match the measured HPBW, i.e. the Airy
pattern for a 10.7~m antenna is used for the 12~m antennas.  An
improved representation of the beam patterns from celestial holography
measurements is currently under test.

When considering the effect of radio frequency interference (RFI), it
is important to know the response in the far sidelobes, which has been
measured on the VLA and VLBA antennas at $\lambda = 18$~cm
\citep{Dhawan2002}.  The declining envelope of the sidelobe response
is consistent with the reference radiation pattern for large diameter
($D/\lambda \geq 100$) parabolic Cassegrain antennas tabulated in
Recommendation SA.509-3 of the radiocommunication sector of the
International Telecommunications Union \citep{ITU2013}.
In general, the gain relative to the main beam will drop below -60~dB
somewhere between $10^\circ-20^\circ$ off axis.

\subsection{Antenna efficiencies}  

For antennas that have a well defined physical collecting area, such
as reflector, lens or horn antennas, the ratio of the effective area
$A_0$ to the physical area $A$ of the aperture is called the aperture
efficiency $\eta$, a dimensionless quantity less than unity:
\begin{equation}
\eta=A_0/A\,.
\label{eq:5}
\end{equation}
The antenna aperture efficiency directly impacts the sensitivity of
the synthesis telescope and is the product of a number of different
loss factors,
\begin{equation}
\eta=\eta_{\rm Ruze}\,\eta_{\rm bl}\,\eta_{\rm s}\,\eta_{\rm t}\,\eta_{\rm r}\,
\eta_{\rm misc}\,,
\label{eq:9}
\end{equation}
where $\eta_{\rm Ruze}=$ reflector surface efficiency, $\eta_{\rm
  bl}=$ reflector blockage efficiency (feed legs and subreflector),
$\eta_{\rm s} =$ feed spillover efficiency, $\eta_{\rm t} =$
illumination taper efficiency, $\eta_{\rm r} =$ panel reflection
efficiency, and $\eta_{\rm misc}=$ miscellaneous efficiency losses due
to reflector diffraction, feed position errors, and polarization
efficiency.  As we will see, the term that is the most frequency
dependent is $\eta_{\rm Ruze}$.  Hence, it is often the case that
observatory documentation will define the aperture efficiency as
$\eta_{\rm 0}\eta_{\rm Ruze}$ where $\eta_{\rm 0}$ is simply the
product of all the other efficiencies.  For example, the ALMA
technical
handbook\footnote{\url{https://almascience.nrao.edu/documents-and-tools/cycle4/alma-technical-handbook}}
quotes $\eta_{\rm 0} = 0.72$.  It is important to realize that $\eta$
is typically elevation-dependent, with the best values occurring at
moderate elevations (usually between 45-60\deg). Gain curves showing
$\eta$ vs. elevation for the VLA antennas are stored in CASA
(Figure~\ref{gaincurves}) and can be used to correct for this varying
amplitude response in the data.

\articlefiguretwo{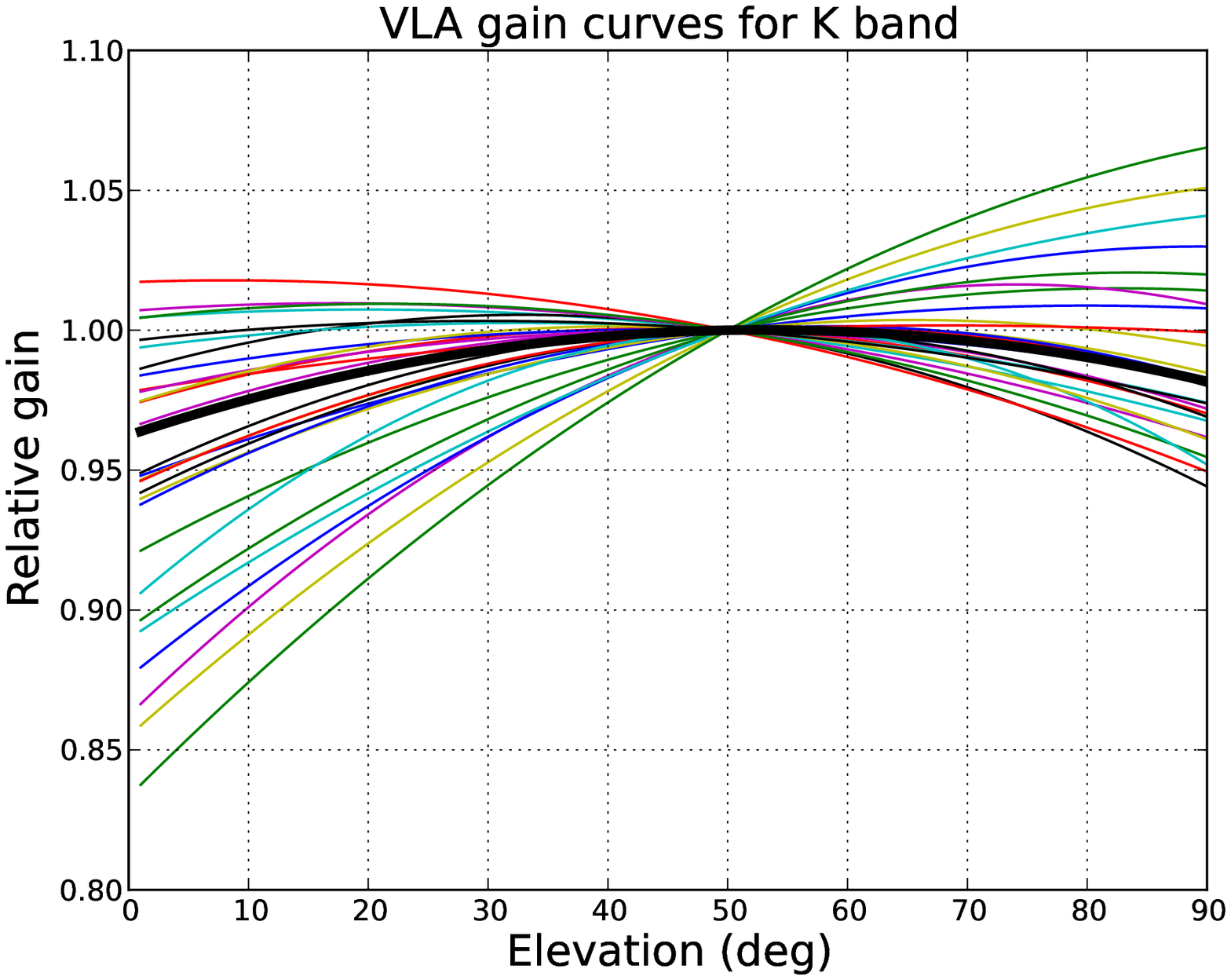}{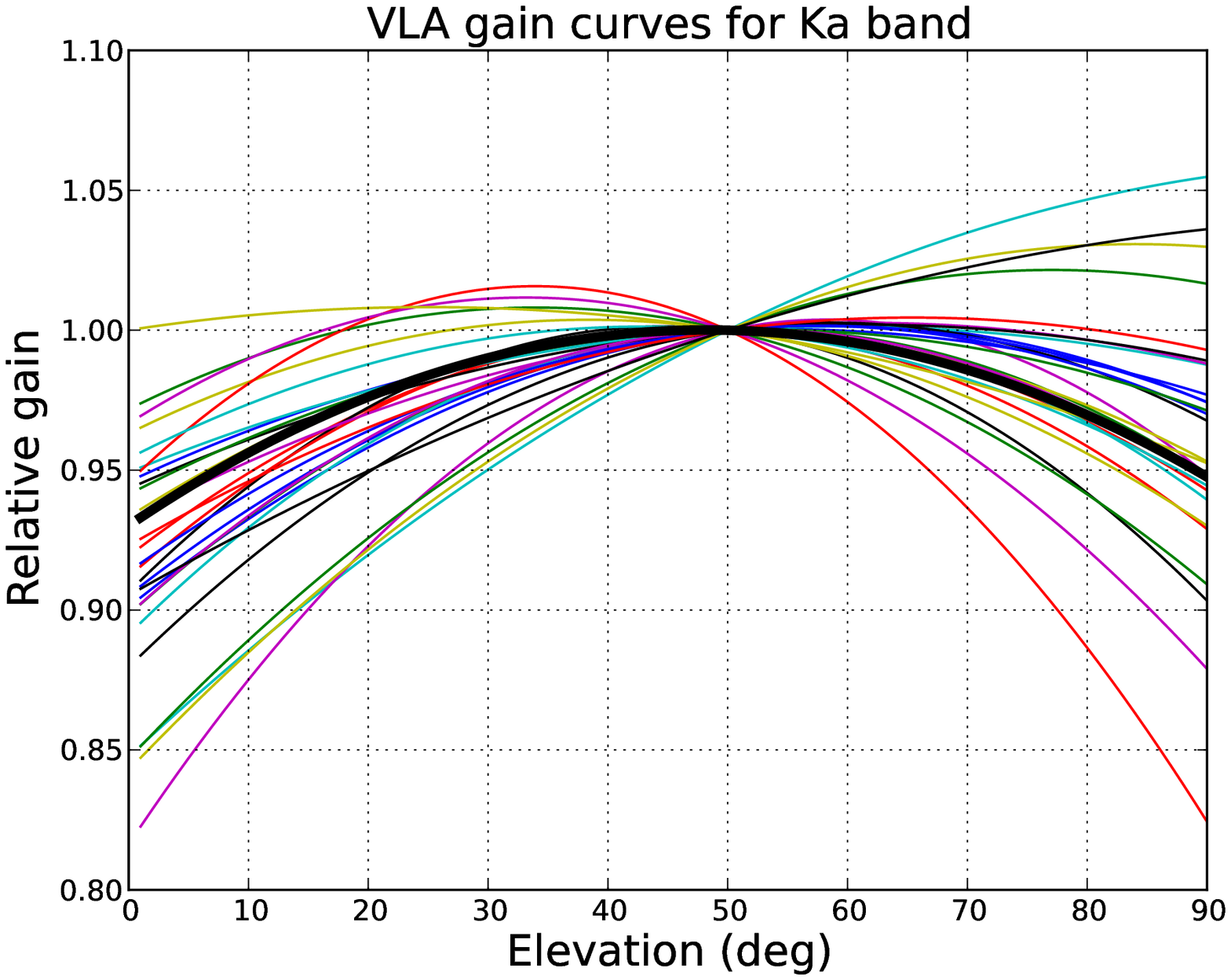}{gaincurves}{Gain
  curves stored in CASA for all the VLA K and K$_{\rm a}$-band
  receivers as of December 2014 (all antennas overlaid). In each plot,
  the thick line is the mean curve.  }

\subsubsection{Surface efficiency}

$\eta_{\rm Ruze}$ accounts for loss due to inaccuracies in the profile
of the reflector.  Surface errors cause the electric field from
different parts of the aperture to not add together perfectly in
phase at the feed, leading to a decrease in received power.
\citet{Ruze66} gives an expression for surface efficiency
\begin{equation}
\eta_{\rm Ruze}= e^{-(4\pi \sigma /\lambda)^2}\,,
\label{eq:10}
\end{equation}
where $\sigma$ is the rms surface error, with the errors assumed to be 
Gaussian random and uncorrelated across the aperture.  In a Cassegrain
(or more complicated) mirror system, $\sigma$ is an appropriately
defined composite rms error of the primary and secondary reflector
surfaces, which should always dominate over subsequent (more accurate)
smaller mirrors.  If the errors are correlated over significant
fractions of the aperture, then additional terms are required on the
right hand side of Eq.~\ref{eq:10}, or more accurately, an integration
of the surface profile map must be performed.  Eq.~\ref{eq:10}
predicts that for an rms error of $\lambda /16$, $\eta_{\rm
  Ruze}=0.54$, which is often taken to define the useful upper
frequency limit ($\nu_{\rm upper}$) for a reflector. For the VLA, with
$\sigma \approx 400\mu$m, $\lambda / 16$ corresponds to $\nu_{\rm
  upper}=47$~GHz. For ALMA, with $\sigma \approx 20\mu$m, $\nu_{\rm
  upper}=940$~GHz.  Most of the drop in $\eta_{\rm Ruze}$ occurs
between 0.5-1.0$\nu_{\rm upper}$, as can be seen in
Table~\ref{rxtable}.

As well as the loss of sensitivity resulting from a low value of
$\eta_{\rm Ruze}$, one must be concerned with the quality of the
primary beam.  The surface errors cause scattering which produces a
broad pedestal surrounding the main lobe of the beam that can be
higher than the usual diffraction-limited sidelobes. This pedestal can
enhance image artifacts caused by sources near the primary beam. For a
reflector of diameter $D$, if the reflector errors are correlated over
distances $D/N$ then the scatter pattern will be $N$ times broader
than the diffraction-limited main lobe, and often correspond to the
panel segment size.  Measurements of this pattern can be made by
scanning large objects like the Moon \citep{Schwab07,Greve98}.  Good
$\eta_{\rm Ruze}$ performance requires careful structural design for
wind, thermal and gravitational loading, together with precise
reflector panels \citep[e.g.][]{Bosma98,Ezawa00} and an accurate panel
setting technique (see \S~\ref{holography}).

\subsubsection{Blockage efficiency}

 The feed or subreflector and its multi-legged support structure block
 the aperture of a reflector antenna.  This typically results in a
 blockage efficiency in the range $0.75<\eta_{\rm bl}<0.95$. A formula
 for $\eta_{\rm bl}$ is given \citep{Lamb86} by
\begin{equation}
\eta_{\rm bl}=\left(1-\frac {\rm \ effective \ blocked \ area}
{\rm total \ area}\right)^2\,.
\label{eq:11}
\end{equation}
The effective blocked area is the blocked area weighted for the
illumination taper in the aperture \citep[see
  also][]{Goldsmith02}. Similarly, the total area is weighted for the
illumination taper in the aperture.  Equation~\ref{eq:11} shows, for
small blockage, that the loss in efficiency is twice the fractional
blocked area. As well as the loss in aperture efficiency, the increase
in antenna beam sidelobe level due to blockage is important for
synthesis telescopes.  Using the Fourier transform relationship, the
form of the antenna voltage pattern with blockage can be calculated as
the unblocked voltage pattern minus the voltage patterns of the
blocked areas.  As a practical example, the ALMA 12~m antennas are of
three different designs (Vertex, AEC and Melco, corresponding to
the three funding partners: North America, Europe, East Asia), and the
effect of their different blockage can be seen in their respective
beam patterns.  The feedleg design of the AEC antennas is
significantly different from the other two designs in that the four
struts are mounted entirely from the edge of the dish
(Figure~\ref{beams}).  In contrast, the feed struts of the Vertex and
Melco antennas meet the dish in several places along the outer half of
the dish, meaning that scattering occurs twice--once on the way from
the sky down to the primary mirror (plane wave scattering) and again
on the way back up to the subreflector (spherical wave scattering).
As shown in Figure~\ref{beams}, the first sidelobe is lower and more
azimuthally uniform on the AEC antennas compared to the Vertex
antennas.

\articlefigure{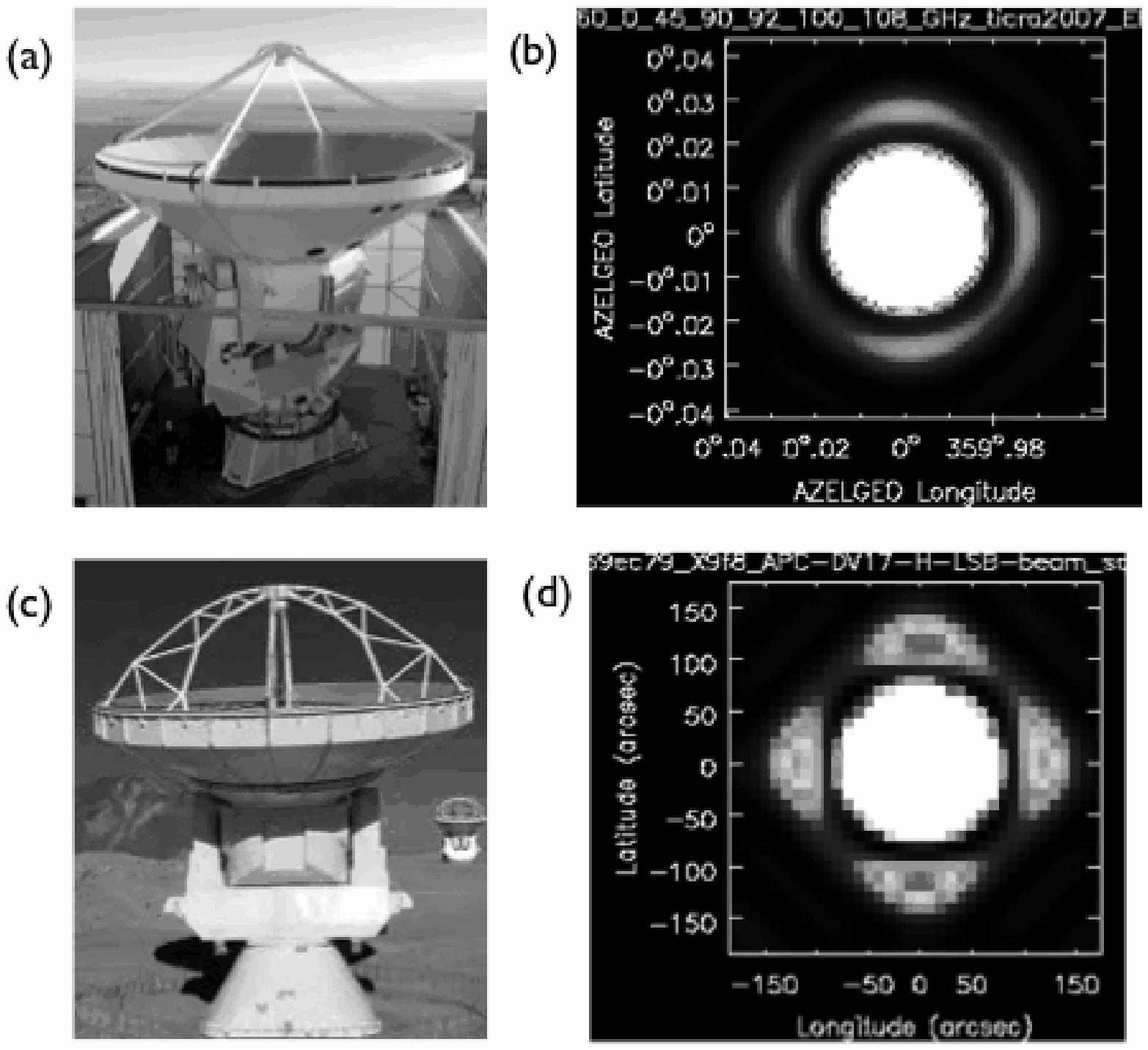}{beams}{(a) Photograph of a 12~m ALMA
  AEC antenna whose feedlegs block only the plane wave from the sky;
  (b) 100~GHz amplitude beam pattern of an AEC antenna obtained from
  celestial holography; (c) Photograph of a 12~m ALMA Vertex antenna
  whose feedlegs also block the spherical wave (i.e., between the
  primary and secondary mirror); (d) beam pattern of a Vertex antenna.
  Note the differences in the first sidelobe due to the difference in
  feed leg geometry.  }

\subsubsection{Feed spillover and illumination taper efficiency}

These two efficiency terms are related to one another, and their
product is sometimes (confusingly) referred to as the illumination
efficiency.  The spillover efficiency can most easily be understood by
considering the antenna in transmission, rather than reception mode.
The spillover efficiency is the fraction of the power radiated by the
feed that is intercepted by the subreflector for a Cassegrain feed, or
by the main reflector for a prime focus system.  Clearly, power that
does not intercept the reflector is lost, and we can be confident that
a similar loss occurs in reception mode by invoking the Reciprocity
Principle \citep[][p.~11]{Rudge82}.  Simultaneously, the illumination
taper efficiency arises whenever the outer parts of the antenna are
illuminated at a lower power level than the central portion, and hence
contribute lower ``weight'' in the aggregate signal (similar to
applying a uv-taper in synthesis imaging).  The spillover and taper
efficiencies can be computed using the integral formulas given in
\citet{Napier99}; but in a qualitative sense, it should be obvious
that adjusting the taper in one direction will generally improve one
term at the expense of the other.  For unshaped, classical Cassegrain
systems (like ALMA antennas), the illumination taper that gives the
best trade-off \citep[i.e., -10~dB,][]{Goldsmith87} will produce a
spillover efficiency of $\approx0.9$, a taper efficiency of
$\approx0.9$, and consequently, a net product of $\approx0.8$.  By
comparison, for the VLA antennas, whose illumination pattern is much
closer to uniform, the net product is $\approx0.9$ \citep{Napier83}.

\subsubsection{Panel reflection efficiency}

Aside from surface errors encompassed by the $\eta_{\rm Ruze}$ term,
smooth unpainted aluminum surfaces generally have a very high
reflectivity at centimeter through submillimeter wavelengths
\citep[typically $\geq0.99$ per mirror,][]{Ezawa00}.  Addition of
paint, which provides long-term protection, adds a small amount of
loss at centimeter frequencies due to additional scattering \citep[up
to a few percent,][]{Lamb92}.  However, above 100~GHz the dissipative
loss of the paint's dielectric material becomes significant, which is
why the panels of most (sub)millimeter telescopes are left unpainted.
Although unpainted, ALMA panels are slightly roughened in order to
scatter infrared radiation to enable safe observations of the Sun.

\subsubsection{Miscellaneous efficiencies}

Not included in the previous efficiency terms is the effect of
diffraction at each aperture.  Whenever the focusing mirror diameters
are large compared to the wavelength of observation, diffraction
losses are low (a few percent or less).  However, these losses become
significant at the long wavelength end of many telescopes.  For
example, at 1~GHz, the diffraction efficiency of the VLA antennas is
0.85 \citep{Napier83}.  

\label{misceff}

The ideal amplitude response of the primary beam is circularly
symmetric with respect to the optical axis, with constant phase out to
the first null, and alternating by $180\deg$ in successive sidelobes.
In reality, small errors in alignment of the subreflector with respect
to the primary surface (i.e. focus errors) can produce a non-circular
beam, which is accompanied by a reduction in efficiency
\citep{Butler03,Goldsmith02}.  Furthermore, any small errors in the
alignment of the receiver feed with respect to the optical axis
(termed an ``illumination offset'') produce non-uniform phase response
in the outer portion of the main beam \citep{Holdaway01}.  In addition
to loss of efficiency, this effect can produce problems when imaging
extended objects and may require special calibration if the
misalignment is significant \citep{Bhatnagar08}.

\subsection{Surface setting techniques}

\label{holography}

Precision multi-panel reflectors are composed of individual panel
segments typically with four or five screw adjustment points per
panel.  The initial setting of the surface segments is generally
performed with a mechanical alignment device or theodolite-assisted
technique, which can achieve accuracies of $\lesssim1$ part in $10^5$
of the total aperture diameter $D$. Further refinement is often done
using photogrammetry \citep{Kesteven12,Feng10,Miller03}.  This technique entails
placing reflective tape targets on the corners of each panel, imaging
the entire surface with a high resolution digital camera from various
angles, and solving for the best fit surface profile. Applying manual
surface adjustments based on this measured profile can typically reach
accuracies of a few parts in $10^6$ of $D$.  The ultimate surface
performance is usually achieved using microwave holography, a
technique developed during the 1970's \citep{Napier71,Bennett76} and
in use at nearly every radio through submillimeter observatory
\citep[e.g.][]{Hunter11,Baars07,Baars99,Grahl86}.  In this context,
the term holography refers to the process of mapping the complex
(amplitude and phase) beam pattern of an antenna and Fourier
transforming the data to the aperture plane.  The angular extent of
the map (typically $1-2\deg$) determines the linear resolution on the
dish.  The phase map provides the surface deviations in units of the
observed wavelength; thus, the wavelength sets the ultimate accuracy,
but it is constrained by the available sources of radiation.
Centimeter-wave telescopes typically use geosynchronous broadcast
satellite signals in $K_u$ band ($\lambda=2.5$~cm), either their
analog continuous wave beacons (i.e. spectral line observed with a narrowband filter) or their broadband digital transmissions.  Continuum from bright quasars can also be used \citep{Kesteven93,Padin87}.
Millimeter-wave telescopes need higher precision (hence higher
frequency sources) and typically use ground-based 3~mm transmitters
mounted on towers, though wavelengths as short as 0.4~mm have
been used \citep{Sridharan02}.

With a well-tuned holography system, the measurement errors are
typically about 0.5-1 part in $10^6$ of $D$.  In addition to measuring
panel misalignment, holography can be used to measure illumination
offsets (\S~\ref{misceff}), systematic antenna panel mold error
\citep{Hunter11} as well as large-scale deformations due to thermal
effects.  Large-scale features can also be measured by celestial
holography, in which bright maser lines \citep{Morris88} or quasar
continuum emissions are used as the radiation source.  A sidereal
source provides the advantage that dish deformations can be measured
at different elevations.  All of these forms of traditional holography
require a second stationary antenna as a reference signal.  An
alternative technique called ``phase retrieval'' or ``out-of-focus''
holography can be performed on a single antenna by mapping the
amplitude beam and fitting for the associated phase error
\citep{Nikolic07}.

\subsection{Pointing model and metrology}

Each antenna in a synthesis array employs a pointing model which
continuously converts the requested topocentric coordinates into the
actual encoder coordinates that will put the antenna on-source.
Pointing models account for basic effects such as encoder zero
offsets, collimation error, non-perpendicularity of the axes, and pad
tilt, as well as higher-order terms that account for gravitational
flexure, encoder eccentricity and other mechanical asymmetries
\citep[e.g.][]{Mangum06,Patel04}.  Some antennas employ additional
metrology, such as thermometers \citep[GBT,][]{Prestage09} and tilt
meters and linear displacement sensors \citep[ALMA,][]{Rampini12}, in
order to input real-time dynamic terms.  The pointing model must be
fit to ``all-sky'' pointing datasets, typically consisting of
continuum scans on dozens of quasars scattered all about the sky,
using fitting software such as TPOINT \citep{Wallace94}.  When an
antenna is relocated to a different pad, at least several terms of the
pointing model must be remeasured and updated to counteract the small
but inevitable changes in geometry.  With the best pointing model in
place, the so-called ``blind pointing'' performance refers to the
typical pointing error after slewing to a particular direction.  For
the VLA, the blind pointing error is typically $10''$ rms in calm
nighttime conditions, but can exceed $60''$ during the day.  To
improve the pointing accuracy during higher frequency observations,
the technique of offset pointing is employed, in which local pointing
corrections are periodically measured toward a bright quasar within
$\sim10\deg$ of the science target.  This technique can reduce the rms
error to $\approx3-5''$ for VLA antennas, and $\approx1''$ for ALMA
antennas (see Fig.~\ref{tracking}).
\label{pointing}

\subsection{Tracking errors and servo control system}

\label{tracking}

In addition to static pointing error, antenna servo tracking errors
are always present at some level and can cause time variation in the
visibility amplitudes, particularly at high frequencies where the
primary beam is smallest.  At the VLA, the rms tracking error is about
$3''$ in low wind, which yields typical peak excursions of $\pm6''$,
or 10\% of the primary beam at 43~GHz.  While this sounds relatively
harmless, it becomes critical when imaging extended objects.  For
example, at the phase center of a Gaussian beam, an rms tracking error
of 10\% of the beamwidth yields an amplitude variation of only 2.7\%;
however, at the half-power point of the beam, it yields a variation of
$\pm28$\% (Figure~\ref{trackingFigure}).  Tracking accuracy is
generally the worst in high or gusty winds.  In benign weather
conditions, the tracking accuracy of an antenna is ultimately set by
the quality of its servo control system.  A description of a modern
digital antenna servo control system for the 6~m diameter SMA antennas
is given in \citet{Hunter13}.  Good servo systems are designed with
safety as the highest priority, with interlocking emergency-stop and
hardware limit switches connected directly to the power source.
Additional software safety measures include: 1) software motion
limits, 2) consistency checks between the position feedback devices
(encoders) and integrals of the velocity feedback devices
(tachometers), 3) monitoring of motor currents and temperatures, 4)
routine servicing of a watchdog timer which will trigger a system
shutdown in case of a processing hangup, and 5) re-engaging of
mechanical brakes whenever power is lost.

\articlefiguretwo{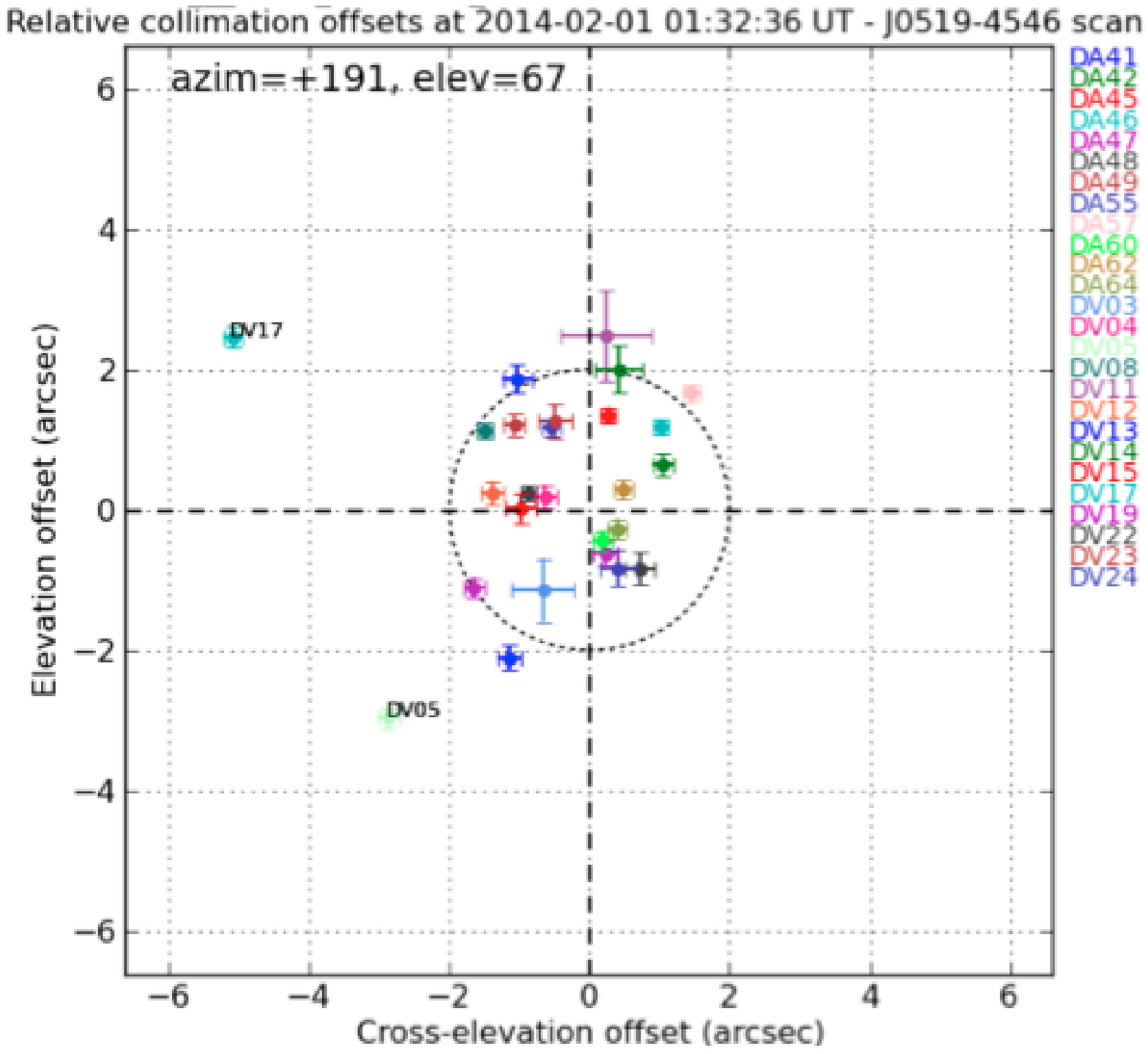}{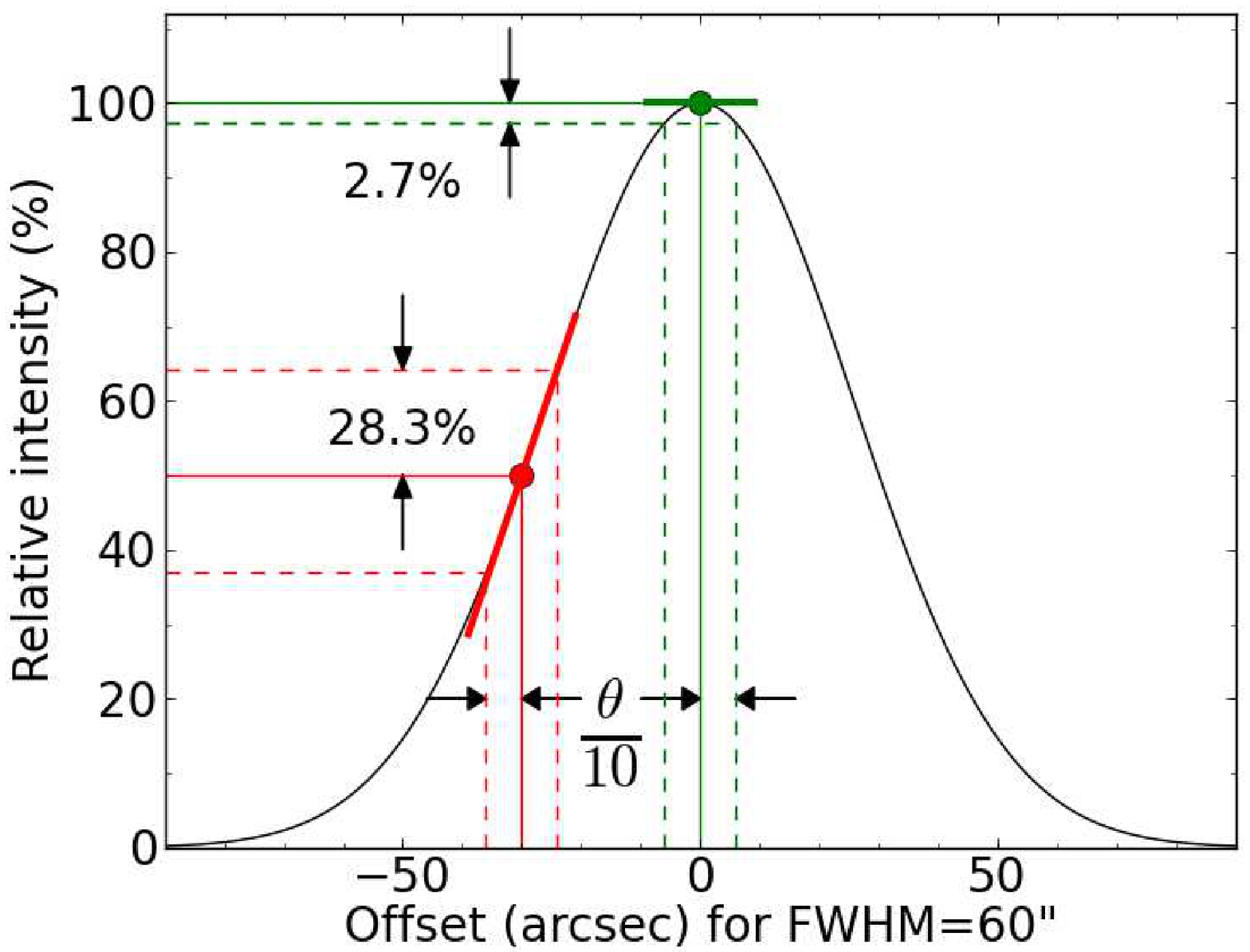}{trackingFigure}{Left)
  Example of the blind pointing performance of ALMA in Band 3. The antenna
  offsets have been measured in an interferometric pointing scan and
  are mostly within the $2''$ specification.  Right) Comparison of the
  amplitude variation expected for a source at the center of a
  Gaussian beam vs. a source at the half-power point, in the face of
  only a modest tracking error (rms = 10\% of the HPBW).} 

While all of the safety logic is running relentlessly in the
background, the system must also compute the instantaneous torque to
apply to track a celestial source at the sidereal rate, or perhaps
perform faster on-the-fly imaging \citep{Mangum07}.  A typical servo
design implements nested loops in which the calculations and
adjustments operate at the appropriate rate.  For example, the
azimuth/elevation position loop runs at $\approx10$~Hz and computes
velocity commands to send to the azimuth/elevation velocity loop
running at $\approx100$~Hz, which in turn commands the motor current
(torque) loop at $\approx10$~kHz.  In the past, the gains of these
loops (traditionally consisting of proportional, integral and
derivative terms) were set in hardware by fixed values of resistors
and capacitors. Similarly, the time constants of any filters on the
feedback signals had to be set in this manner.  In modern systems,
these gains and filters are configurable in software, and can be
adjusted (if necessary) when operating in different modes or under
different conditions.  Also, command-shaping can be implemented to
smoothly transition between slewing and tracking modes to prevent
overshoot and more rapidly acquire the target \citep{Hunter13}.
Finally, more complicated algorithms can be attempted to try to
achieve faster response times while avoiding the excitation of
structural resonances \citep{Gawronski08}. Both the GBT
\citep{Whiteis12} and VLA \citep{Jackson11} are in the midst of an
upgrade of their original servo system hardware and software.
  
\section{Receivers}

The role of the receivers in a synthesis telescope is to linearly
amplify weak radio frequency signals while adding minimal noise, and
down-convert them into room-temperature analog output signals on
coaxial cables at intermediate frequencies (IFs) suitable for
digitization.  Receivers are traditionally comprised of a {\it
  front-end} (FE) and a {\it back-end} (BE).  The FE includes the
components that {\it must} be attached to the antenna, in contrast to
the BE which includes the electronics (sometimes called the {\it IF
  chain}) that can be mounted in a separate rack and which process the
IF signals.  In this chapter, we will cover the FE polarization
splitting device, the FE detector, and the BE electronics preceding
the digitizers.

\subsection{Overview of receiver technology}

\subsubsection{General configuration}

Three current technological limitations can essentially explain the
configuration of receivers in radio and (sub)millimeter astronomy.
First, we can build broadband low-noise amplifiers (LNAs) with optimal
noise performance up to about 120~GHz \citep[see the review
  of][]{Pospieszalski05}. Second, we can digitize signals of
instantaneous bandwidth up to about 2~GHz, which require a Nyquist
sampling rate \citep{Nyquist28} of 4 gigasamples/second
(Gs/s).\footnote{In fact, higher speed samplers are now being
  developed and fielded at submillimeter observatories
  \citep[e.g.][]{Patel14}, but 4~Gs/s was the limiting bandwidth when
  ALMA and EVLA were being designed.}  From these two facts, we can
see that to observe at $\nu_{\rm RF} \gtrsim 120$~GHz, the first
device in the front-end must be a device to downconvert the signal
(called a mixer) instead of an amplifier.  Also, to observe with an
aggregate bandwidth $>2$~GHz, a mixer must be present in the IF chain,
regardless of the observing frequency.  The third limitation is that
when used as the first component in the front-end, both mixers and
amplifiers must be cooled to cryogenic temperatures to yield
competitive performance\footnote{An exception to this rule is at low
  frequencies ($\nu_{\rm RF}<$400~MHz) where the Galactic background
  increases from tens to thousands of Kelvin.  In this regime, the
  room-temperature performance of LNAs is adequate without introducing
  extra noise.}.  LNAs reach their optimal performance at about 15~K,
which requires only a two-stage cryostat.  In contrast, the best
submillimeter mixers are superconducting tunnel junctions which
require a more complicated three-stage cryostat to reach their optimal
operating temperature of $\leq4$~K, a practical disadvantage compared
to LNAs at $\nu_{\rm RF} \lesssim\/120$~GHz.  Putting together these
facts, we can surmise that most ALMA FEs must begin with a cold mixer
followed by a cold LNA, while VLA FEs begin with a cold LNA.
Following these components, both ALMA and VLA receivers require room
temperature mixers and amplifiers in their BEs, prior to digitization.

\subsubsection{Polarization separation}

For optimal sensitivity, we want to build dual-polarization receivers
that can accept both polarizations from astronomical targets.  This
ability requires either dual linear or dual circular feeds.  Because
FE amplifiers and mixers operate on individual polarization signals, a
polarization splitting device is needed.  There are two broad
categories of devices that can provide $\sim20$~dB of polarization
purity with low loss: waveguide and quasioptical.  The most common
waveguide devices are called ortho-mode transducers (OMTs) and are
placed between the feed horn and the FE device.  An OMT is a four-port
microwave device with three physical ports.  It accepts a dual
polarization signal into its one common input port (typically a square
or circular waveguide) and splits the two polarizations into separate
physical output ports (either rectangler waveguide or coax).  OMTs can
be designed numerically using software that solves Maxwell's
equations.  Although it can be challenging to achieve octave-wide
(operating over factor of 2 in frequency) designs with good isolation
and low loss \citep{Skinner91}, they are fairly straightforward to
machine when the dimensions are large, i.e. at wavelengths longward of
1~mm \citep[e.g.][]{Asayama09,Kamikura10}, but good results have been
obtained at 0.6~mm \citep[ALMA Band~8;][]{Naruse09}.  The most common
quasioptical splitter is a wire grid which reflects one polarization
and transmits the other.  Wire grids must be placed at a beam waist
preceding the feedhorn, and the required wire separation is
$\lesssim\lambda/20$ for good performance
\citep{Sorenson10,Houde01,Chambers88}.  At high frequencies, wire
grids are easier to construct than OMTs.  The disadvantage of wire
grids is that you need two feedhorns (one per polarization) instead of
one and each has an accompanying mirror to refocus the beam after the
grid.  Thus, optical alignment can be tricky and often leads to a
significant {\it beam squint}, a condition in which the primary beams
of the two polarizations differ in pointing direction by $\sim$0.1
beam (the ALMA specification) or worse.

\subsection{VLA receivers}

The VLA receiver system \citep{Perley09} consists of 10 bands (see
Table~\ref{rxtable} and Figure~\ref{rxphoto}). The two lowest
frequency bands employ crossed dipole feeds in front of the
subreflector followed by room temperature amplifiers.  The rest of the
bands employ offset-Cassegrain corrugated feed horns followed by
cryogenic LNAs housed in individual cryostats.  The polarization
splitters \citep{Coutts11a,Coutts11b} in between the feedhorns and the
LNAs consist of quadruple-ridge OMTs ($\nu<12$~GHz), waveguide
(B\o{ifot} junction) OMTs ($12<\nu<40$~GHz), or sloped septums
(Q-band).  At the input of the LNAs, most of the bands (those above
4~GHz) also employ isolators, which are passive, non-reciprocal
two-port devices which, like a subway turnstile, prevent the
propagation of signal in the reverse direction.  In this case, the
isolators serve to reduce leakage between the polarizations and reduce
standing waves in the optics.

\articlefigure{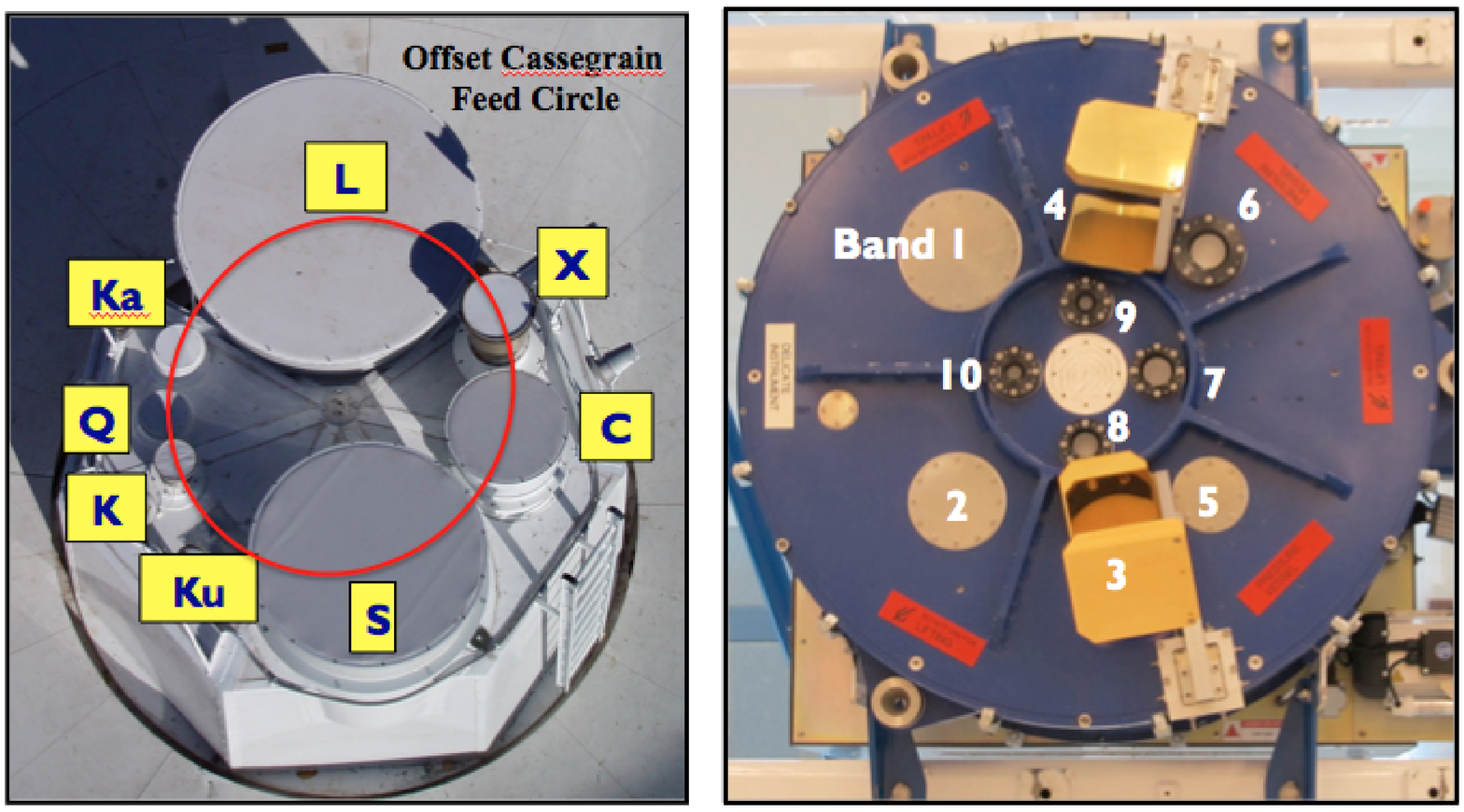}{rxphoto}{Left) Photograph of the VLA's
  offset Cassegrain feed circle, as viewed from the subreflector;
  Right) Top view of the ALMA cryostat with the vacuum windows labeled
  by band number (Table ~\ref{rxtable}).  The room temperature optics
  for Bands 3 and 4 are in place above their respective windows.}

\begin{table}[h]
\begin{center}
\caption{VLA and ALMA Receiver bands and their properties \label{rxtable}}
\small
\begin {tabular}{ccccccc} \hline\hline
\noalign{\vspace{3pt}}
\multicolumn{2}{c}{Receiver band} & Frequency range & $\eta_{\rm Ruze}$\tablenotemark{a} & \multicolumn{2}{c}{Polarization} & Sideband\\  
Central $\lambda$ & code          &   (GHz)         &                 & type & splitter & type \\
\noalign{\vspace{3pt}} \hline \multicolumn{7}{c}{VLA receivers}\\
\tableline \tableline \noalign{\vspace{2pt}} 4 m & 4 & 0.058--0.084 &
1.0 & dual linear & dipole & SSB \\ 90 cm & P & 0.23--0.47 & 1.0 &
dual linear & dipole & SSB \\ 20 cm & L & 1--2 & 1.0 & dual circular &
q.-r. OMT & SSB \\ 13 cm & S & 2--4 & 1.0 & dual circular & q.-r. OMT &
SSB \\ 6 cm & C & 4--8 & 0.99 & dual circular & q.-r. OMT & SSB \\ 3 cm
& X & 8--12 & 0.97 & dual circular & q.-r. OMT & SSB \\ 2 cm & K$_{\rm
u}$ & 12--18 & 0.94 & dual circular & wg. OMT & SSB \\ 1.3 cm & K &
18--26.5 & 0.87 & dual circular & wg. OMT & SSB \\ 1 cm & K$_{\rm a}$
& 26.5--40 & 0.74 & dual circular & wg. OMT & SSB \\ 7 mm & Q & 40--50
& 0.57 & dual circular & septum & SSB \\ \tableline
\multicolumn{7}{c}{ALMA receivers}\\ \tableline \tableline 7 mm & 1
(Q) & 35--51\tablenotemark{b} & 1.0 & dual linear &
OMT\tablenotemark{b} & SSB \\ 4 mm & 2 (E) & 67--90\tablenotemark{b} &
1.0 & dual linear & OMT\tablenotemark{b} & SSB \\ 3 mm & 3 (W) &
84--116 & 0.99 & dual linear & OMT & 2SB \\ 2 mm & 4 & 125--163 & 0.99
& dual linear & OMT & 2SB \\ -- & WVR & 175.3--191.3 & 0.98 & single
linear & -- & DSB \\ 1.6 mm & 5 & 163--211 & 0.98 & dual linear & OMT
& 2SB \\ 1.3 mm & 6 & 211--275 & 0.96 & dual linear & OMT & 2SB \\ 0.9
mm & 7 & 275--373 & 0.93 & dual linear & wire grid & 2SB \\ 0.7 mm & 8
& 373--500 & 0.87 & dual linear & OMT & 2SB \\ 0.45 mm & 9 & 600--720
& 0.74 & dual linear & wire grid & DSB \\ 0.35 mm & 10 & 787--950 &
0.59 & dual linear & wire grid & DSB \\ \noalign{\smallskip}
\tableline \multicolumn{7}{l}{$^{a}$Assuming $\sigma= 400\mu$m for VLA
and $\sigma=20 \mu$m for ALMA}\\ \multicolumn{7}{l}{$^{b}$Bands under
development}\\ \multicolumn{7}{l}{References for ALMA bands: 1: \citet{Huang16};
 3: \citet{Kerr14,Claude14};}\\
\multicolumn{7}{l}{4: \citet{Asayama08}; 5: \citet{Billade12}; 6: \citet{Kerr14}; }\\
\multicolumn{7}{l}{7: \citet{Mahieu12}; 8: \citet{Sekimoto08}; 9: \citet{Baryshev15};}\\ 
\multicolumn{7}{l}{10: \citet{Gonzalez14,Fujii13,Uzawa13}, WVR: \citet{Emrich09}}\\
\end{tabular}
\end{center}
\end{table}

\subsubsection{Amplifiers}

An amplifier is an active, two-port device, meaning that it requires a
voltage supply, and has one RF input and one RF output.  The input
signal emerges at the output with greater power (typically 10--30~dB,
i.e.  10x--1000x) and is unchanged in frequency.  The current
generation of NRAO cryogenic LNAs on the VLA, VLBA, GBT, and
other telescopes employ heterostructure field effect transistors
(HFET) which operate at 15~K.  They deliver a noise temperature
performance of $\sim4$~K at low frequency and about 5 times the
quantum noise limit at high frequency
\citep[$\nu$>12~GHz,][]{Pospieszalski12}:
\begin{equation}
T_{\rm LNA}(\nu>12\text{GHz}) \approx 5h\nu/k = 10(\nu/42\text{GHz})~~\text{K}.
\end{equation}
These indium-phosphide (InP) HFETs come from the ``cryo3'' series of
wafers manufactured by Northrup Grumman Space Technology in 1999.  At
each frequency range, there exists an optimal size (gate periphery)
for a transistor that facilitates the design of a broadband amplifier.
The gate peripheries range from 200~$\mu$m at Ku-band (and below) down
to 30~$\mu$m at W-band.  The HFET devices are housed inside a block of
gold-plated brass and the input is connected by either coaxial cable
(low frequency receivers) or waveguide (high frequency receivers) to
the corresponding OMT output.  The LNA package has a total gain of
$\approx 35$~dB.  To avoid feedback, the HFETs are mounted in a
channel small enough to block all waveguide modes in the band of
operation. Electrical connections are made via microstrip and bond
wires.  It is important to note that the LNA is not the only
contributor to the overall receiver temperature ($T_{\rm rx}$).  While
the details and measurements can be found in the EVLA memo series, a
rough model for the mid-band $T_{\rm rx}$ for VLA bands above 12~GHz
is $T_{\rm rx} \approx 2 + 0.5\nu_{\rm GHz}$~K.

\subsubsection{Mixers}
\label{mixers}
Mixers were invented around the time of World War I for radio
direction finding \citep[see the historical review of][]{Maas13}.  A
mixer is a three-port device which accepts two inputs: a broad RF
signal and a narrow LO continuous wave (CW) tone, and produces a
broadband IF output signal. This frequency conversion occurs in a
diode with a strongly nonlinear current ($I$)-voltage ($V$)
characteristic, $I = f(V)$. As an example, consider a simple
square-law diode for which
\begin{equation}
\label{nonlinear}
I = \alpha V^2.
\end{equation}
When the two input fields are superposed:
\begin{equation}
\label{added}
V = V_{\rm RF}\cos(\omega_{\rm RF}t) + V_{\rm LO}\cos(\omega_{\rm LO}t)
\end{equation}
the non-linear nature of the mixer will effectively multiply the LO
and RF signals together.  This effect can be seen by substituting
equation Eq.~\ref{added} into Eq.~\ref{nonlinear}, which produces
three terms for the resulting current, including the cross term which
can be expanded by a trigonometric identity into:
\begin{equation}
2V_{\rm RF}V_{\rm LO}\cos(\omega_{\rm RF}t)\cos(\omega_{\rm LO}t) = V_{\rm RF}V_{\rm LO}[\cos(\omega_{\rm RF}-\omega_{\rm LO})t+\cos(\omega_{\rm RF}+\omega_{\rm LO})t].
\end{equation}
Thus, current will flow at the difference frequency of the LO and IF.
It is important to notice that the multiplication serves to transfer
the phase from the RF signal to the IF signal, a process called
heterodyning.  Because they can transfer
phase in this manner, mixers are key components for interferometers.

Mixers range from low-cost, off-the-shelf packages using Si Schottky
diodes that operate at room temperature (in cell phones etc.)  to the
expensive, delicate, research-grade
Superconductor-Insulator-Superconductor (SIS) tunnel junctions that
were developed in the late 1970s \citep{Dolan79,Richards79,Rudner79}
and early 1980s \citep{Pan83,Wengler85} and today serve as the
cryogenic FE detector for ALMA Bands 3 through 10. At terahertz
frequencies, the performance of SIS mixers declines and hot-electron
bolometer mixers are used instead \citep[e.g.][]{Meledin04}.  Room
temperature mixers are also employed in ALMA and VLA IF circuitry
after the received signal has been sufficiently amplified so that
their conversion loss is of little consequence.  Typically, the IF
signal output by a mixer will contain overlapping signals from two
frequency ranges termed {\it sidebands} (see Figure~\ref{mixer}),
resulting in double sideband (DSB) performance.
\articlefigure{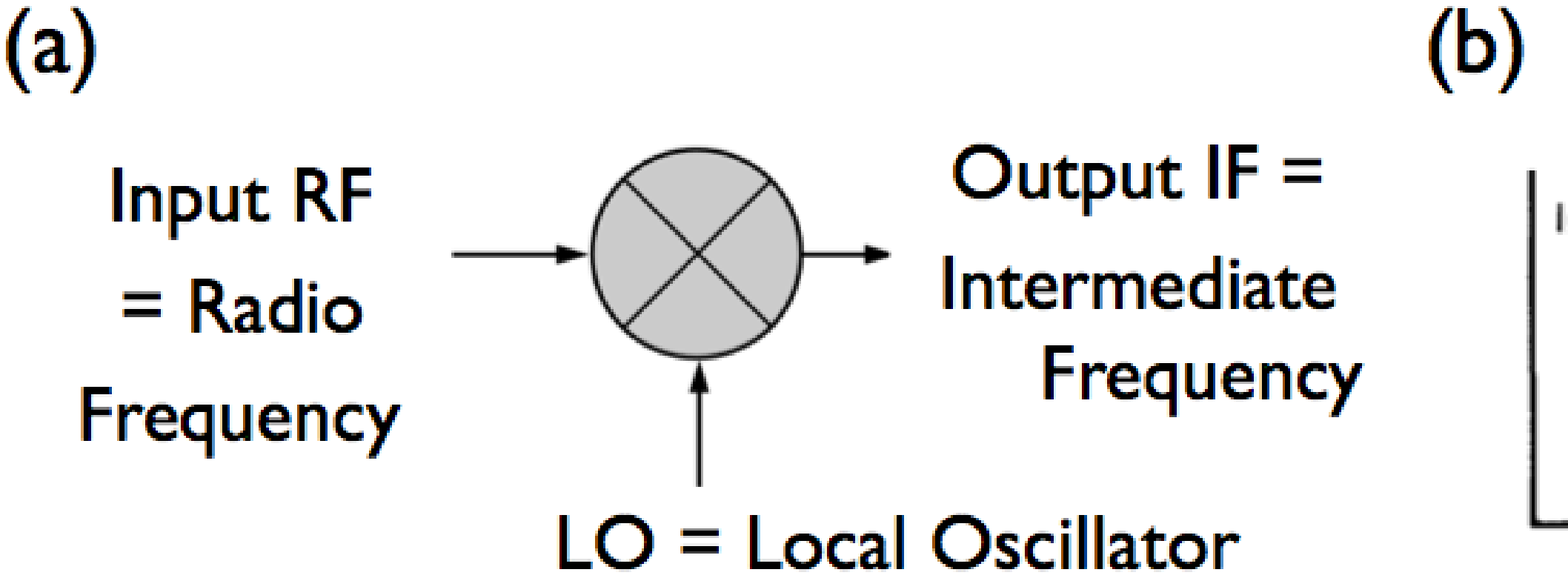}{mixer}{a) Schematic representation of
a mixer's input and output signals; b) Spectrum showing the relative
frequencies and bandwidths of the signals.}
One sideband represents a piece of the RF spectrum above the LO
frequency and the other is a piece below the LO frequency.  The DSB
confusion can be avoided either by pre-filtering the RF signal (termed
an SSB mixer), or by designing a mixer to separate the sidebands into
separate IF outputs (termed a 2SB mixer).  If only one of the
sidebands from a 2SB mixer is desired, the IF output corresponding to
the unwanted sideband can be terminated into a load (such as a
waveguide absorber), which is a configuration termed a 1SB mixer
\footnote{Another way to achieve a 1SB mixer is to terminate the
  unwanted RF sideband of a 2SB mixer into a cold load ahead of the
  mixer using a Martin-Puplett diplexer \citep{Martin70,Lesurf88}.}.
VLA receivers employ SSB mixers, as is likely for ALMA bands 1 and 2
(still under development).  ALMA bands 3 through 8 are 2SB, while
bands 9 and 10 are DSB, as are all receivers on the SMA
\citep{Blundell04}.  The relative sensitivities of these mixer types
has been evaluated for the ALMA site \citep{Iguchi05,Jewell97}, but in
general 2SB is superior to DSB.  A development project to upgrade the
band 9 mixers into 2SB format is underway \citep{Khudchenko12}.

\subsection{ALMA receivers}

The ALMA receiver system consists of ten frequency bands
(Table~\ref{rxtable}), all housed in the same 0.97~m diameter, 450~kg
cryostat (Figure~\ref{rxphoto}) which is mounted inside the receiver
cabin at the Cassegrain focus.  Currently, a receiver band is brought
into focus by adjusting only the pointing of the primary mirror to
orient the desired sky direction toward the desired receiver window.
There is some improvement to be gained from also adjusting the
subreflector to tilt approximately half-way toward the selected
receiver \citep{Hills05}, but this added complication has not yet been
introduced into the system since the pointing and focus models will
need to account for it.  A low-loss polymer membrane, initially made
of Gore-Tex \citep{Candotti06,Koller06} but replaced by Teflon
fluorinated ethylene propylene (FEP), protects the cabin from the
outdoor environment.  The ALMA FE optics have three generic layouts
\citep{Rudolf07}.  Bands 1 and 2 use room-temperature (external to the
cryostat) polymer lenses as the focusing element.  Bands 3 and 4 are
in the outermost position in the cryostat and use a pair of external
room-temperature mirrors \citep[Band 3 includes a Teflon lens on the
corrugated feedhorn,][]{Claude08}.  The focusing optics for the higher
bands are off-axis ellipsoidal mirrors mounted on the cold cartridge
assemblies (CCAs) inside the cryostat.  The diameters of the cryostat
windows are set large enough to avoid significant truncation losses
\citep{Lamb03}.  For polarization separation, ALMA bands 7, 9 and 10
use wire grids while the other bands use OMTs (Table~\ref{rxtable}).
The SIS mixers and their enclosing CCAs in the ALMA receivers were
constructed by different international parties (see the references in
Table~\ref{rxtable}).  In all cases, the receiver noise performance
meets specification, beginning with 41~K in Band 3, and essentially
following the function $4h\nu/k$ in the higher bands.  Finally, the
room-temperature Dicke-switched water vapor radiometer (WVR) in each
antenna has proven effective and essential to removing atmospheric
phase fluctuations on short timescales, down to the adopted
integration time of 1.152~second ($6\times0.192$), and on all baseline
lengths \citep{Nikolic13}.

\subsection{Local oscillators (LOs)}

Being required to drive mixers (\S~\ref{mixers}), an LO signal must be
a clean tone with high signal to noise ratio (SNR) in order to obtain
accurate astronomical spectra.  LOs are constructed from a base
oscillator with a high-$Q$ electro-mechanical feature, such as a
tunable cavity or a resonant sphere, which is ultimately synchronized
to an atomic frequency standard.  For ALMA, the fundamental
oscillators located in each warm cartridge assembly (WCA) are
commercially-produced yttrium iron garnet (YIG) spheres embedded in a
magnetic field generated by the sum of a coarse tuning coil and a fine
tuning (FM) coil.  These compact YIG packages produce clean tones in
the 2-40~GHz range and are also used in the VLA.  For the higher
frequency bands of ALMA, the YIG tone must be multiplied by one or
more integer multiplication stages, many of which include power
amplifiers to boost the multiplied signal.  All of the LOs for ALMA
\citep{Bryerton13} and VLA were built by NRAO.  The LO supplying the
first mixer in the FE is often called LO1 in order to distinguish it
from the LOs in the BE, which are numbered starting from LO2.

\subsubsection{Phase-lock loops (PLLs)}

Although a free-running YIG tone is typically very clean, it is
subject to drift in frequency with time, and its close-in phase noise
(i.e. the line-broadening of the tone) is not negligible.  In
contrast, a radio telescope requires a precisely stable LO frequency
in order to observe spectral line features.  An interferometer has a
further requirement that the receivers in all antennas be phase-locked
so that celestial signals will correlate.  A circuit to stabilize and
lock the LO is called a phase-lock loop (PLL).  A good description of
a modern digital PLL used on the SMA interferometer is given in
\citet{HunterPLL}.  Similar in concept to an antenna servo
(\S~\ref{tracking}), the PLL circuit continuously analyzes the phase
difference with respect to an accurate low-frequency reference signal
(of order 20--100~MHz), which is produced by a device called the First
LO Offset Generator (FLOOG) in NRAO terminology.  The PLL computes and
applies a correction to the FM tuning magnetic coil of the YIG in
order to maintain lock.  The bandwidth of the correction circuit is
typically a 0.5-1~MHz, which enables rapid re-locking after a Walsh
cycle phase change (see \S~\ref{Walsh}).  Initial lock is achieved by
starting from a computed (or a look-up table) tuning value and
sweeping the coarse coil until the tone is at the prescribed location
in the IF of the PLL \citep[see, e.g.][]{Garcia12}.  A PLL typically
relies on an external mixer to downconvert the signal from the LO
being controlled to a value close to the frequency of the
low-frequency reference, and this external mixer in turn relies on an
accurate high-frequency reference signal for its LO.  For ALMA, this
reference signal is delivered by a photomixer \citep{Huggard02}
located in each WCA which converts the photonic frequency reference
distributed via fiber optic cable and originating from the laser
synthesizer \citep{Ayotte10} in the Central LO \citep{Shillue12}.  In
fact, the fundamental references for LO2 in the BE and for the FLOOG
in the WCA are also distributed on the same fiber using wavelength
division multiplexing.

\subsubsection{LO modulation}

\label{Walsh}

In an interferometer, aside from enabling the mixers to operate, the
LOs are also crucial for implementing many additional features that
ensure data quality \citep{Thompson07}.  Very fine control of the
frequency and phase of LO1 is inserted via the PLL reference generated
by a direct digital synthesizer (DDS), which is part of the FLOOG in
NRAO systems.  For example, suppression of spurious tones and DC
offset is achieved by modulating LO1 with $180\deg$ phase switching
using a Walsh function sequence \citep{Granlund78,Emerson08} and
removing it after digitization by adjusting the digital signal
datastream in a supplementary fashion (i.e. via a sign change).  This
technique serves to wash out any signals in the digital datastream
that did not enter at the FE mixer (i.e. those that did not originate
from the sky).  In high spectral resolution modes, $180\deg$ phase
switching does not work well \citep{Napier07}, so a complementary
technique is used by ALMA called LO offsetting \citep{Kamazaki12}.  In
this case, each antenna's LO1 is shifted by a different small amount
(in integer steps of 30.5 kHz = 15625/512), then this shift is removed
downstream in LO2 or a combination of LO2 and the tunable filter bank
(TFB) LO (sometimes called LO4) in the baseline correlator.  In the
VLA, the LO offsetting technique is called ``f-shift'' and uses prime
number frequency steps, which are removed by the fringe rotators in
the Wideband Interferometer Digital Architecture (WIDAR) correlator
\citep{Carlson03}.  A spurious signal that enters the system between
the insertion and removal points receives a residual fringe frequency
equal to the spacing of the shifts between antennas and is suppressed
upon normal integration of the data.  In particular, LO offsetting can
supply more than 20 dB of additional rejection of the unwanted image
sideband in 2SB receivers. This extra suppression is crucial to
eliminate strong spectral lines whose remnants may otherwise survive
the moderate (10-20 dB) suppression supplied by a 2SB receiver
alone. In practice on ALMA, LO offsetting also reduces residual
closure errors.  Finally, the two sidebands in a DSB receiver
(\S~\ref{mixers}) can be quite effectively separated ($\geq20$~dB) by
applying $90\deg$ Walsh phase switching.  In this manner, both
sidebands can be recovered simultaneously in the correlator, thus
doubling the effective bandwidth, as is done on the SMA
\citep{Patel14}.  However, Walsh switching will not remove image
signals which are not common to all antennas, including the
atmospheric noise from the image sideband in a DSB receiver.

\subsection{Back-end components}

\subsubsection{Round trip phase (RTP) measurement and correction}

The required distribution of the LO reference signals in large arrays
like ALMA and VLA occurs via many kilometers of fiber optic cables.
Optical fibers offer many advantages to radio interferometers,
including low loss and wide bandwidths \citep{Young91}.  However, the
thermal expansion coefficient of these fibers combined with the
diurnal temperature change of the environment leads to temporal
changes in the optical path length of the fiber that vary as a
function of antenna pad location.  Although the cables are buried, the
thermal effect is still significant and causes delay changes to the LO
references carried by the fiber\footnote{The SMA uses custom-designed
  low thermal coefficient optical fiber that is no longer commercially
  produced. Over the modest fiber lengths (500m or less), it obviates
  the need for an RTP system.}.  Mechanical stresses on the
above-ground portions of the fiber are also significant
\citep{Daddario98}.  To compensate for these problems, ALMA employs a
line length correction (LLC) system, which uses a piezoelectric fiber
stretcher driven by a PLL to maintain a constant optical path length
on each fiber during an observation \citep{Shillue12}.  In principle,
the stretcher requires only enough range to compensate drift between
visits to the phase calibrator, provided that the stretchers are only
ever reset to the center of their range in between consecutive
integrations on the phase calibrator {\it and} that the offline
calibration software (CASA) is aware of the resets.  In practice, this
synchronization has not been implemented in the ALMA system (nor in
CASA), mainly because the stretchers have been found to have
sufficient range to handle drifts over a few hours (even on 10~km
baselines) before slipping a fringe, which is longer than current
(Cycle 3) standard ALMA observing blocks ($<90$~minutes).

The EVLA project also built a round trip phase (RTP) measuring system
to mitigate project risk when it was not clear what the stability
performance of the fiber would be \citep{Morris08,Durand02}. The RTP
was designed to be able to measure and send corrections to the EVLA
WIDAR correlator.  Although the VLA has longer baselines, it operates
at lower frequencies, and the amount of phase change in the fiber
observed by the RTP was small and slow compared to atmospheric phase
variations (and thus are effectively removed by the normal phase
calibration sequence). However, the RTP was very useful in diagnosing
phase vs. elevation changes due to antenna electronics and temperature
changes {\it not} in the fiber, thus allowing those problems to be
identified and fixed.  Although the RTP system was deactivated in
2010, it can be re-activated if needed, e.g. for a future Pie Town
link.

Although it is not implemented as a round-trip measurement, the VLBA
has a pulse-cal system to measure the relative instrumental phase
between baseband channels \citep{Walker95,Daddario96}.  A train of
1~MHz or 5~MHz pulses from a tunnel-diode are injected into a
directional coupler, the same one used to inject the signal from the
$T_{\rm cal}$ noise diode (see \S~\ref{tcal}).  The pulses pass
through the LNA and all downstream processing and the resulting phases
are detected for selectable tones in the back-end. These so-called
$P_{\rm cal}$ data are supplied to the user and can be used to
calibrate the phase characteristics (offset and frequency slope) of
the antenna's components independent of the atmosphere or whether a
calibrator is being observed.


\subsubsection{Square law detectors and IF level setting}

Two unglamorous but essential parts of the receiver BE are the ability
to: 1) measure the total power of signals using square law detectors
(SQLDs) which convert the power in a continuum signal into voltage
\citep{Bare65}; and 2) adjust signal levels in order to optimize the
inputs into successive devices along the IF chain.  For example, when
the input signal to an amplifier exceeds its specified input range, it
no longer functions in a linear fashion. In other words, the effective
gain factor is less than what it would be with a smaller input. This
effect is termed ``gain compression'', and is often accompanied by
other bad characteristics including spurious oscillations and change
in the bandpass shape.  To avoid these pitfalls, all observations
begin with the insertion of a known load (possibly just blank sky)
into the beam followed by a sequential adjustment of the programmable
attenuators placed strategically along the IF chain in order to
achieve optimal levels, which are implemented as target values at each
set of SQLDs.  An attenuator is a wideband two-port device that
dissipates a fraction of the input power into a resistor.
Programmable attenuators typically have a range of 0-15 or 0-31~dB
with steps of 0.5 or 1~dB.  

After the final mixing stage, the signal to be digitized has reached
its lowest frequency, which is traditionally termed a {\it baseband}
even if the low end of the band is not at 0~Hz.  In ALMA and VLA, each
of the 2~GHz-wide basebands covers 2-4~GHz.  After passing through an
analog anti-aliasing filter \citep{Holmes08}, the ALMA baseband
signals are sampled in the second Nyquist zone of the 3-bit 4~Gs/s
sampler \citep{Deschans02}.  Adjusting the power level in each
baseband is also essential as it represents the input level to the
digitizers, which often have a narrow range of input power over which
the SNR is optimal, and is particularly true for ALMA.  In fact, ALMA
observations must set and remember two different sets of attenuation
levels -- one set used for the system temperature scans (to avoid
saturation on the hot load) and one set used for normal observing.
The VLA includes an additional device called a gain slope equalizer,
which allows the signal power to be adjusted in a frequency-dependent
manner providing a more uniform level across the baseband fed to the
digitizers \citep{Morgan07}.  The ALMA Band 10 receiver also includes
an equalizer in the IF section of its WCA \citep{Fujii13}.  Finally,
SQLDs can also be used independently of the autocorrelation portion of
the correlator to perform continuum pointing or focus scans, and to
measure the baseband-averaged system temperature T$_{\rm sys}$.

\subsection{Measurement of system gain and sensitivity during observations}

\subsubsection{$T_{\rm rx}$ and $T_{\rm sys}$ measurement at ALMA}

\label{trxtsys}

In all ALMA observations, the $T_{\rm rx}$ and $T_{\rm sys}$ in each
baseband is measured periodically to be able to properly calibrate the
data and account for atmospheric absorption.  By design, the
measurement is a spectral measurement in order to capture the inherent
variation of the receiver sensitivity across the observing band, and
to capture the spectral variation in atmospheric opacity due to
molecular absorption features \citep[e.g.]{Paine04}.  The measurement
is achieved using three autocorrelation spectra measured sequentially
on: the sky, the ambient temperature load, and the heated load.  The
loads are temperature-controlled blackbodies with nearly perfect
emissivity \citep[0.999,][]{Murk10} mounted in the mechanically-driven
ALMA Calibration Device (Figure~\ref{acd}).
\articlefigure{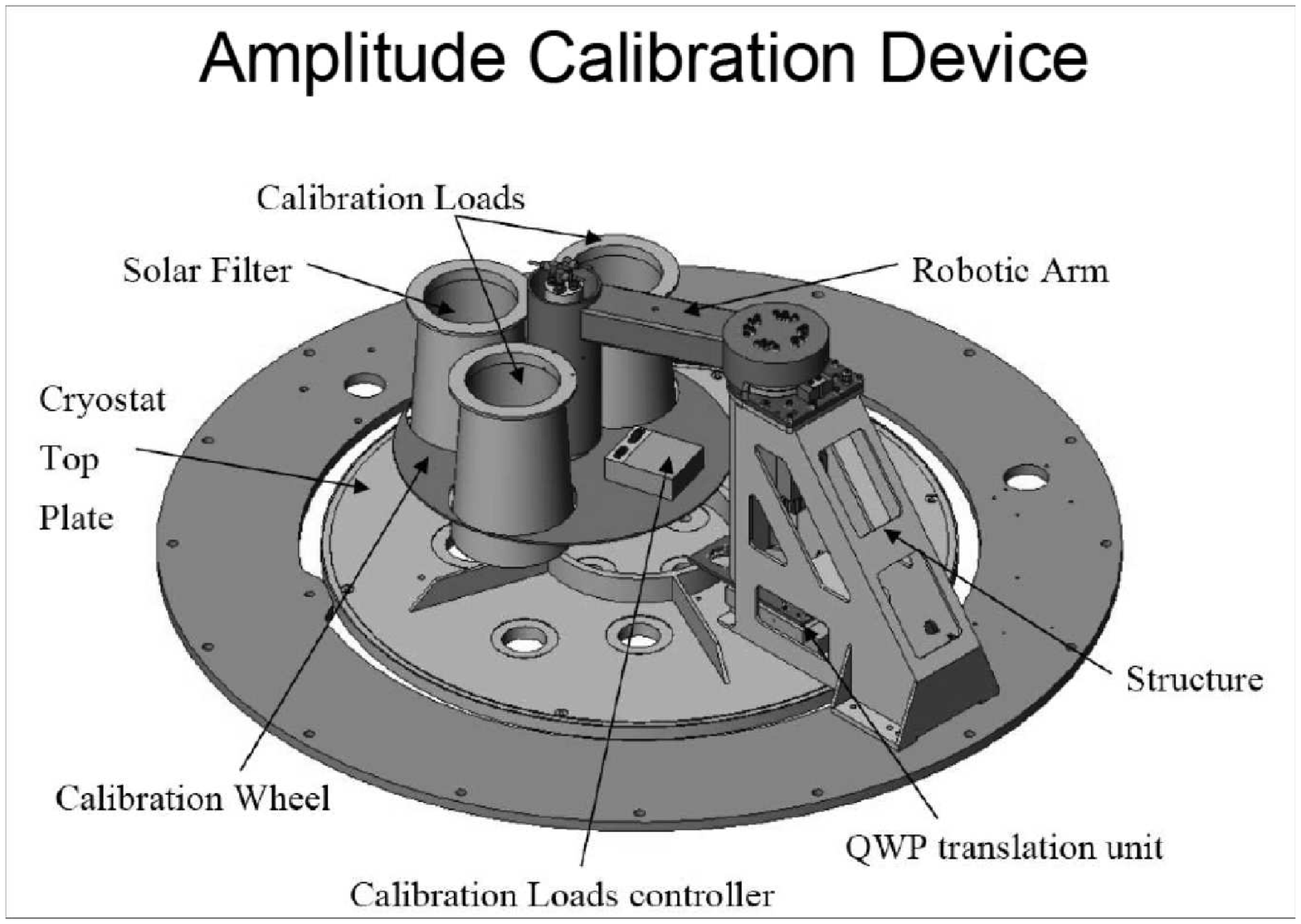}{acd}{Diagram of the ALMA Calibration Device
  located between the cryostat and the membrane, and mounted on the
  Front End Support Structure (the outer ring in the
  diagram). The quarter wave plate (QWP) unit was designed to convert
  the Band~7 linear feeds to circulator polarization, but it is not
  currently populated. }
The online software \citep[TelCal,][]{Broguiere11} takes the two load spectra to compute $T_{\rm rx}$ via the $y$-factor method (see Eq.~12
in the chapter on Basics of Radio Astronomy).  TelCal then uses this result along with the sky spectrum and the atmospheric model
\citep{Pardo01} to compute $T_{\rm sys}$ using the chopper wheel method \citep[e.g.][]{Jewell02}.  It accounts for the relative sideband gain, currently using a single value per baseband, but it could eventually be a spectrum.  These temperature spectra are stored in the data and applied offline in CASA to place the visibilities onto
an absolute temperature scale and to set the relative weights of the data.  An example of the three measurements and the resulting $T_{\rm rx}$ and $T_{\rm sys}$ spectra are shown in Figure~\ref{tsys}.
\articlefigurethree{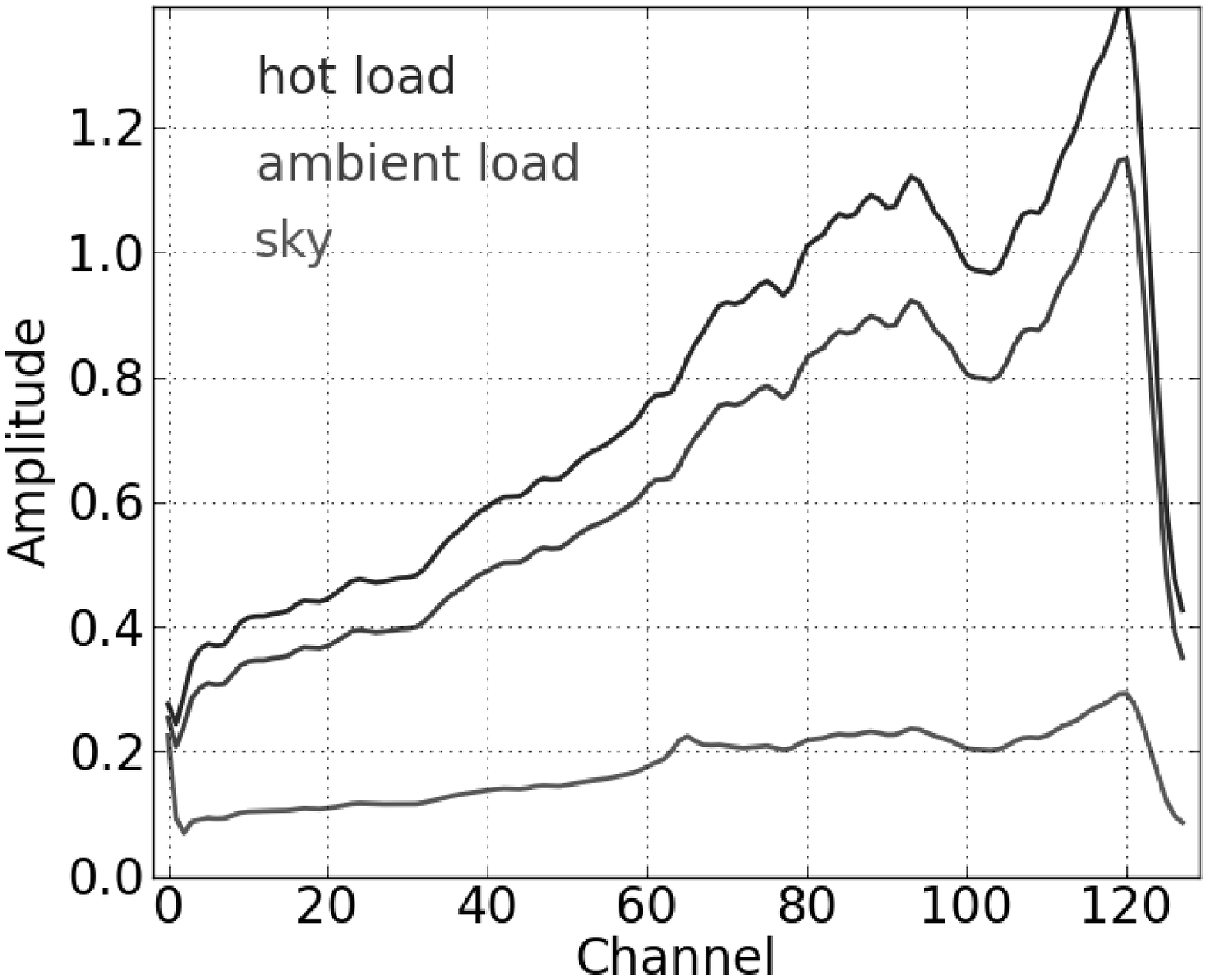}{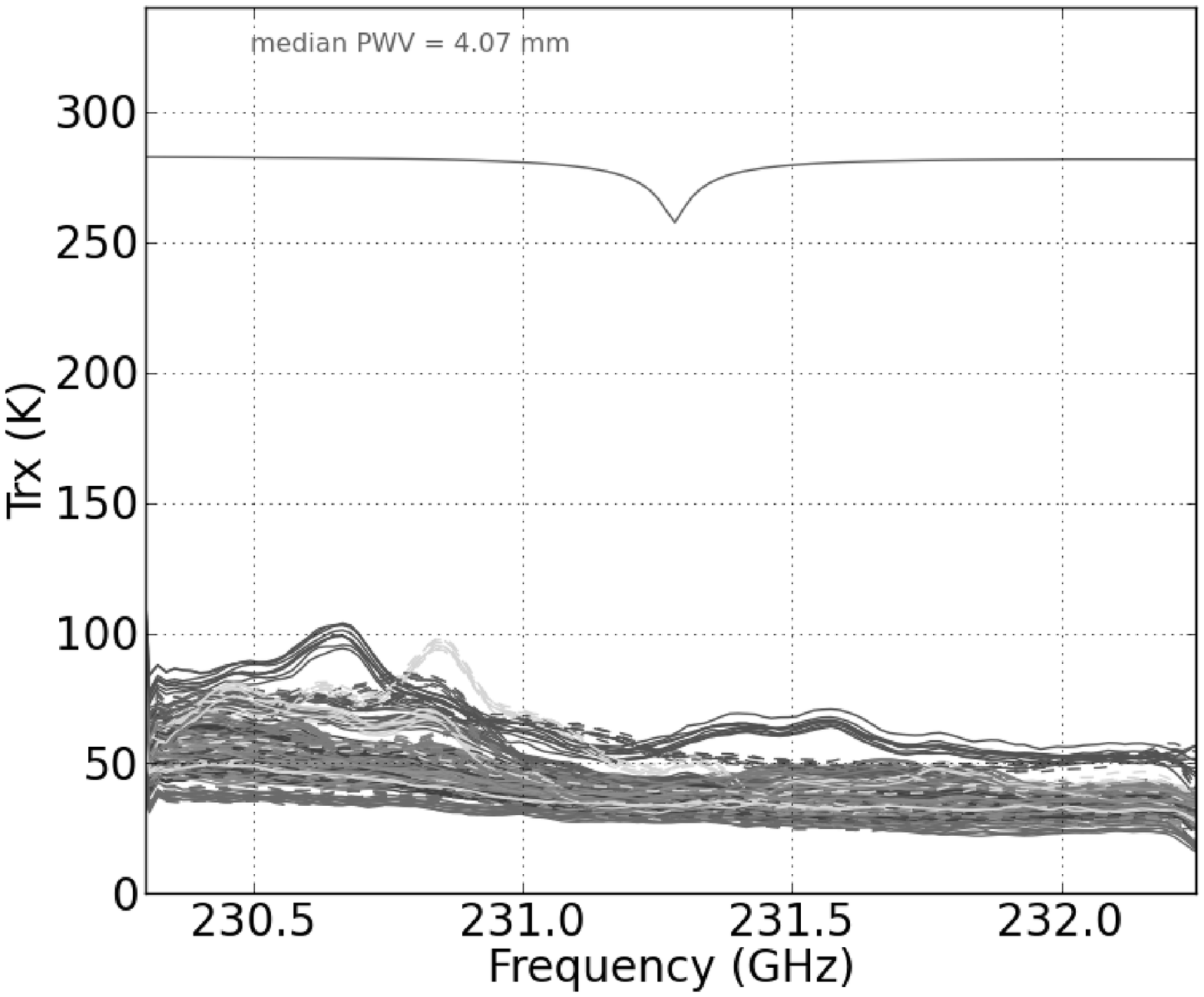}{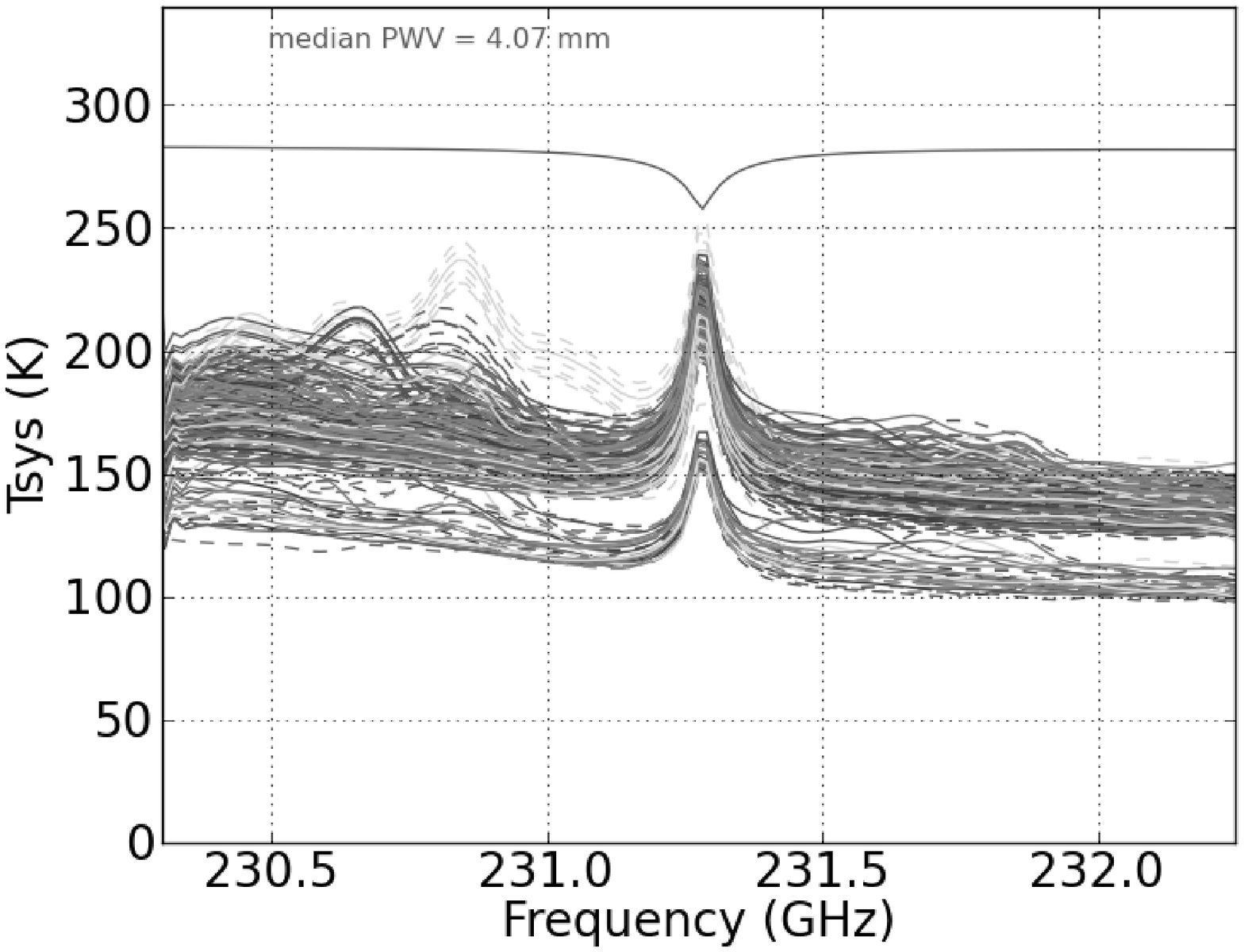}{tsys}{Left panel) An overlay of three successive ALMA observed spectra in Band~6 from one 2~GHz-wide baseband acquired: on the sky, on the ambient load and on hot load; Middle panel) The $T_{\rm rx}$ spectra from all antennas and both polarizations overlaid on one another; Right panel) The corresponding $T_{\rm sys}$ spectra overlaid on the same vertical scale as the $T_{\rm rx}$.  The upper curve in the latter two panels is the atmospheric transmission spectrum, showing the presence of an ozone (O$_3$) line ($16_{1,15}-16_{0,16}$), which causes an increase in the $T_{\rm sys}$ in those channels.}
Although the power level varies by a factor of three across the
baseband, the overall receiver sensitivity is fairly constant.
However, the fact that the $T_{\rm sys}$ spectra show upward spikes at
the frequencies of atmospheric molecular absorption lines (in this
case from ozone) reflects the fact that those channels are less
sensitive in the data, and hence can be down-weighted (using the
spectral weights option available in CASA version $\geq 4.3$).  This
capability is important to obtain the best performance when combining
all the channels of all the spectral windows to obtain a single
multi-frequency synthesis (mfs) continuum image.  Currently, all
$T_{\rm sys}$ spectra are obtained in the low resolution mode of the
baseline correlator, time division mode (TDM), which yields 15.625~MHz
channels in dual-polarization mode \citep{Escoffier07}.  For higher
resolution spectral windows obtained in frequency division mode (FDM),
the $T_{\rm sys}$ correction must be interpolated to narrower channels
when applied to the data in CASA.  A future software upgrade should
allow $T_{\rm sys}$ spectra to be obtained directly in FDM spectral
windows, which will provide an additional benefit of improved removal
of atmospheric lines.  This capability was introduced on the ACA
correlator starting in ALMA Cycle 3.

\subsubsection{$T_{\rm cal}$ and $T'_{\rm sys}$  measurement at VLA}
\label{tcal}
An alternative to a mechanized load system is a broadband calibrated
noise diode that is switched on and off at a high rate
($\approx20$~Hz).  At the expense of a slight increase in system
noise, it provides a stable modulated reference signal.  Each VLA
receiver contains such a diode with a noise temperature ($T_{\rm
cal}$) measured in the laboratory on a 25-100~MHz grid (depending on
the band).  A synchronous detector located in the WIDAR station boards
calculates the sum ($P_{\rm on}+P_{\rm off}$) and difference ($P_{\rm
on}-P_{\rm off}$) powers along with the ratio $R$ and $T'_{\rm sys}$
every second:
\begin{equation}
R = \frac{2(P_{\rm on} - P_{\rm off})}{P_{\rm on} + P_{\rm off}} \text{ and } 
T'_{\rm sys} = \frac{T_{\rm cal}}{R}.
\end{equation}
These values are stored in the switched power table of the
astronomical data.  Application of $T'_{\rm sys}$ in AIPS or CASA
places the visibilities on an absolute temperature scale.  It also
removes any gain variations in the electronics between the diode and
the correlator down to 1~second timescales.  Only a single $T'_{\rm
  sys}$ value per subband is computed, using an interpolated value
from the $T_{\rm cal}$ grid; thus, the spectral resolution of the
correction is typically coarser than ALMA's channelized $T_{\rm sys}$.
Also, in contrast to the $T_{\rm sys}$ discussed in \S~\ref{trxtsys},
the definition of $T'_{\rm sys}$ is with respect to the receiver
input, so it does not account for the effect of the atmosphere.  The
resulting elevation-dependence of the gain can be compensated for in
the offline software using a measurement or estimate of the zenith
opacity.  It is worth noting that WIDAR's digital requantization of
the 3-bit or 8-bit signal into a 4-bit signal introduces an additional
gain change after the synchronous detector.  This gain change is
automatically applied to the 3-bit data online and is also stored in
the same table for reference.

\section{Derivation and simulation of the radiometer equation}

The radiometer equation is fundamental to radio astronomy as it
predicts the expected standard deviation of repeated measurements of
the antenna noise temperature based on finite time samples.  It is
based on a few concepts in physics and statistics which can be easily
simulated on a computer.  Here we present a simulation which
illustrates the fundamental derivation of this equation.  The
derivation can be found in more detail in \citet{Hunter15}.  Shown in
Figure~\ref{gaussian}\/ is the time series and corresponding histogram
of a wide-sense stationary random RF signal (often termed ``white
noise''), whose amplitude ($x_n$) follows Gaussian statistics and has
a mean value of zero and a variance $\sigma^2$ equal to the mean power
$P$ \citep{Shannon49}.
\articlefigure{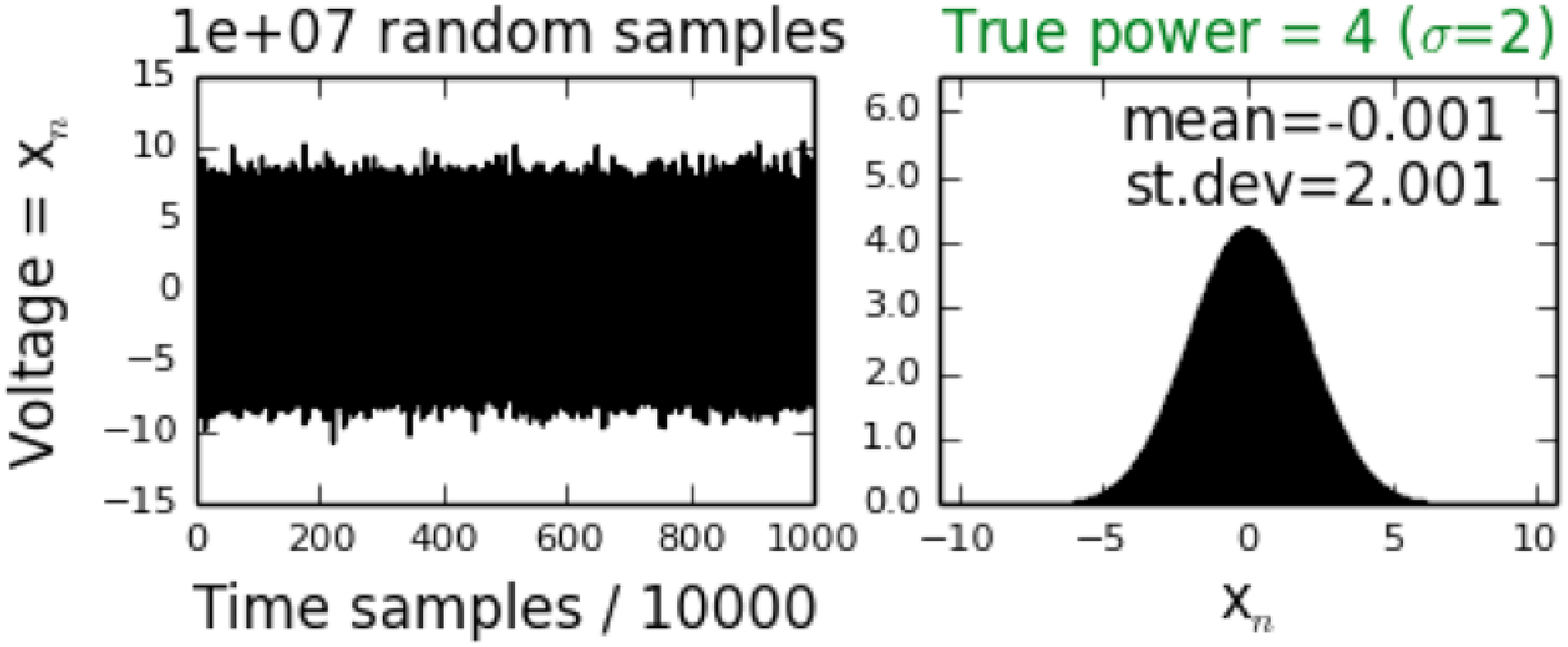}{gaussian}{Simulated time series with
  $10^7$ random samples of a white noise RF signal.  Left panel)
  voltage vs. time.  Right panel) the corresponding histogram of the
  voltage, for which the computed mean and standard deviations are
  listed. }
We have set the mean power to 4 in arbitrary units.  Shown in
Figure~\ref{gaussiantwo} is the square of the amplitude which is the
instantaneous noise power, and its histogram, which follows the expected
gamma distribution.  
\articlefigure{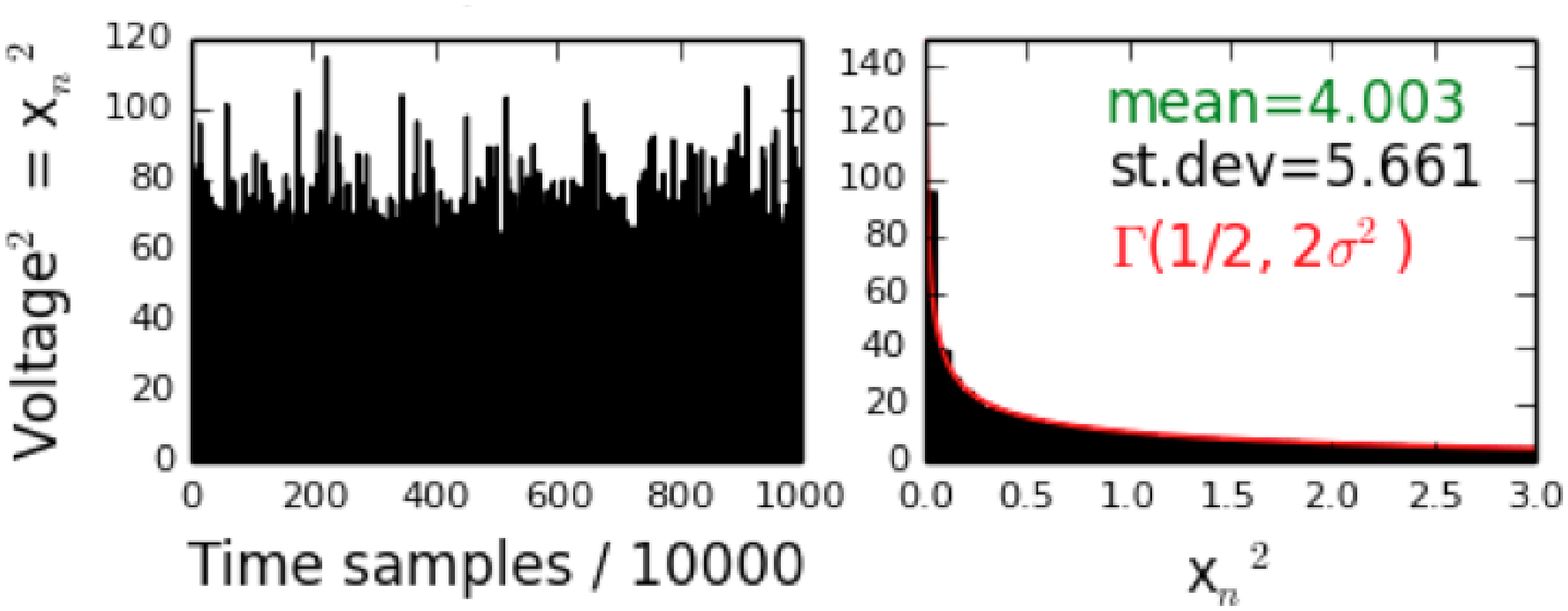}{gaussiantwo}{Left panel) The power vs. time
  of the signal in Figure~\ref{gaussian}. Right panel) the
  corresponding histogram of the power, for which the computed mean and
  standard deviations are listed. The envelope of the histogram is a
  gamma distribution with parameters of 1/2 and 8. }
For any chosen sample of the noise power, statistics tells us that the
standard error of the mean will have a distribution variance of
$\sigma_{\rm P}^2 = 2\sigma^4/N$ where $N$ is the number of samples 
\citep{Fisher30}.
The number of statistically independent samples in time $\tau$ of a
signal with bandwidth ($\beta$) is $N=2\beta\tau$ \citep{Oliver65}.
Thus, the expected standard deviation of the power is $\sigma_{\rm P}
= \sigma^2/\sqrt{\beta\tau}$.  In the limit of large $N$, the mean of
the gamma distribution is the expectation mean of $x_n^2$ and is
equivalent to $P$, thereby yielding the traditional radiometer
equation:
\begin{equation}
\sigma_{\rm P} = \frac{P}{\sqrt{\beta\tau}}.
\end{equation}
Since $P \propto T$ in the Rayleigh-Jean limit \citep{Johnson28}, it
can also be written in terms of $\sigma_{\rm T}$, with $T_{\rm sys}$
in place of $P$ \citep[as is done in Eq.~14 of the chapter on Basics of
Radio Astronomy; see also ][]{Randa08}.

In Figure~\ref{integrate}, we simulate observing the RF signal with a
single-dish telescope for increasing values of the sample size $N$.
In the left panel, we break the datastream of $10^7$ samples into
$10^5$ observations each containing $N=100$ samples.  Each observation
provides an estimate of the noise power and is placed into the
corresponding bin of the histogram.  The distribution peaks near the
value of 4.0 but with a broad uncertainty of $\sigma_{\rm P} = 0.564$,
yielding an SNR of 7.07.  The expected uncertainty of the variance,
$\sqrt{2\sigma^4/N}$, is 0.566.  Thus, the prediction of the
radiometer equation matches the simulated result to $<0.4\%$.  In the
second panel, we increase the size of each observation to 1000
samples, and place each of the resulting $10^4$ noise power estimates
into the corresponding power bin.  The peak of the distribution
remains close to 4 while the width of the distribution becomes
narrower. The uncertainty is now $\pm0.180$ and the SNR is now 22.6.
In this case, the radiometer equation predicts an uncertainty of
$\pm0.179$.  Finally, in the third panel, the size of each observation
is increased to 10000 samples.  The resulting uncertainty in the noise
power is now down to $\pm0.057$, again matching the prediction of the
radiometer equation.  The SNR is now 70.2.  To summarize, we have
increased the observation time $\tau$ by a factor of 100 and the SNR
has improved by a factor of 10.

\articlefigure{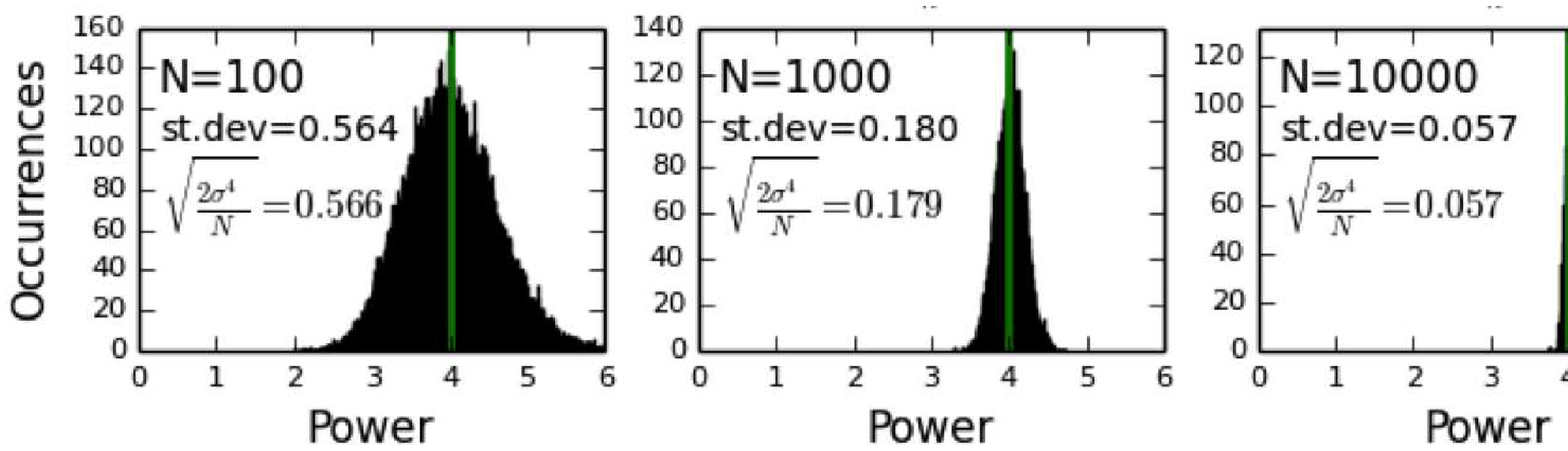}{integrate}{The three panels show the
  results of a computer simulation of the radiometer equation, in
  which $10^7$ random samples of an RF signal with power = 4 are
  successively ``observed'' with three different observation lengths
  containing $N$=100, 1000, and 10000 samples.  The uncertainty of the
  measured power (i.e. its standard deviation) is visualized by the
  width of the histogram, which decreases with $1/\sqrt{N}$.}

The radiometer equation applies equally well to an
interferometer. However, because cross-correlation represents a
multiplication of two independent, zero-mean RF signals (in contrast
to auto-correlation which effectively squares a single signal), the
resulting product is not positive definite like it is in
Figures~\ref{gaussiantwo} and \ref{integrate}.  Instead, the mean of
the distribution gives the correlated power rather than the total
power.  Thus, it can be zero if there is no source in the beam,
although the variance will not be zero.  A similar numerical
simulation shows that the uncertainty of the variance of this
cross-correlation product is: $\sigma_{\rm P} =
\sqrt{\sigma^4/N}$. Thus, the noise on any given baseline is lower
than the single-dish case by is $\sqrt{2}$, which is consistent with
the general formula that the SNR scales as $1/\sqrt{\text{number of
    baselines}}$ \citep[e.g.][]{Wrobel99}. This result implicitly
assumes that the antennas have equal collecting area and efficiency,
and that we are in the limit that the correlated signal is small
compared to the noise.  Of course, the cross-correlation data product
is more commonly examined in terms of amplitude and phase rather than
power.  The expected probability distributions of these component
quantities as a function of the SNR are given in \citet{Crane89}.

We conclude with a note of caution.  All receivers (SIS mixers and
LNAs) exhibit gain fluctuation at some level
\citep[e.g. HEMTs,][]{Gallego04,Wollack95}, which can lead to a
performance that is worse than predicted by the radiometer equation if
the statistics of the noise is non-stationary.  Non-stationary noise
typically exhibits a power law spectrum, i.e. noise $\propto
f^{-\alpha}$, and its variance diverges with time \citep{Hunter15}.
Another source of gain fluctuation is cryostat temperature
fluctuations that reach the cold mixer \citep{Kooi00}, which can be
compensated by adjusting the voltage bias of the subsequent LNA in
real time \citep{Battat05}. Regardless of the origin of the
instability, the receiver total power output stability vs. integration
time is often characterized by the Allan variance \citep[see][and
references therein]{Daddario03}.  At short integrations, the
sensitivity follows the expected curve for white noise but eventually
levels off and will begin to increase at longer integrations.  Thus,
in general, the integration time must be kept short to avoid losing
sensitivity.

\acknowledgements 

The National Radio Astronomy Observatory is a facility of the National
Science Foundation operated under agreement by the Associated
Universities, Inc.  ALMA is a partnership of ESO (representing its
member states), NSF (USA) and NINS (Japan), together with NRC (Canada)
and NSC and ASIAA (Taiwan) and KASI (Republic of Korea), in
cooperation with the Republic of Chile. The Joint ALMA Observatory is
operated by ESO, AUI/NRAO and NAOJ. This research made use of NASA's
Astrophysics Data System Bibliographic Services and the SPIE and IEEE
Xplore digital libraries. The authors thank Vivek~Dhawan,
Anthony~R.~Kerr, and Marian~W.~Pospieszalski of NRAO for comments,
corrections and improvements, and Robert~Kimberk of the
Harvard-Smithsonian Center for Astrophysics for discussions and
research on the radiometer equation.



\end{document}